\documentclass[10pt]{iopart}
\usepackage{iopams,graphicx,lineno,hyperref} 
\usepackage{titlesec}
\usepackage{comment}
\usepackage{multirow}
\usepackage{color}
\usepackage{balance}

\begin{document}

\paper[Deformable microchannels]{Static response of deformable microchannels: A comparative modelling study}

\author{Tanmay C Shidhore and Ivan C Christov}

\address{School of Mechanical Engineering, Purdue University, West Lafayette, IN 47907, USA}

\ead{christov@purdue.edu}

\begin{abstract}
We present a comparative modelling study of fluid--structure interactions in microchannels. Through a mathematical analysis based on plate theory and the lubrication approximation for low-Reynolds-number flow, we derive models for the flow rate--pressure drop relation for long shallow microchannels with both thin and thick deformable top walls. These relations are tested against full three-dimensional two-way-coupled fluid--structure interaction simulations. Three types of microchannels, representing different elasticity regimes and having been experimentally characterized previously, are chosen as benchmarks for our theory and simulations. Good agreement is found in most cases for the predicted, simulated and measured flow rate--pressure drop relationships. The numerical simulations performed allow us to also carefully examine the deformation profile of the top wall of the microchannel in any cross section, showing good agreement with the theory. Specifically, the prediction that span-wise displacement in a long shallow microchannel decouples from the flow-wise deformation is confirmed, and the predicted scaling of the maximum displacement with the hydrodynamic pressure and the various material and geometric parameters is validated.
\end{abstract}

\vspace{2pc}

\submitto{\JPCM}
\maketitle

\ioptwocol

\section{Introduction}
\label{sec:intro}

Microfluidics is a promising field, allowing the miniaturization of various tests, assays and analyses that require the processing of fluids. From analysing DNA and RNA in bodily fluids (e.g., blood) to isolating circulating tumour cells for the purposes of early cancer diagnostics, ``lab-on-a-chip'' devices aim to disrupt the cost and complexity of biomedical laboratory testing. However, ``it can be difficult to translate [these] ideas into the commercial space'' \cite{D17}, as also evidenced by the failure of the Silicon Valley small-sample laboratory testing start-up Theranos \cite{D15}. In part, some of the difficulties are technological, but significant gaps in the basic theory of microfluidic devices also remain. A decade ago, George Whitesides commented that microfluidics is ``in its infancy'' \cite{W06}. Today, even though introductory textbooks on the subject have been written \cite{NW06,B08,K10}, we still lack a complete understanding of some of the fundamental mechanical problems, and microfluidics remains the topic of active research \cite{C10_book,C13_book}.

For example, the relationship between the pressure drop and the flow rate (a key relationship needed for the design and operation of \emph{any} fluidic system) in a deformable microchannel remains a topic of vigorous experimental investigation \cite{GEGJ06,CTS12,OYE13,KRO14,RS16,RDC17}. However, each new study comes with its own theoretical model, which is sometimes of limited applicability. In this work, we aim to provide a general framework and to benchmark results by comparing models against ``full'' three-dimensional, two-way coupled fluid--structure interaction simulations and experimental data from the literature. Specifically, we focus on the static response of deformable microchannels, the building blocks of lab-on-a-chip devices.

Lab-on-a-chip devices are typically manufactured from poly(dimethylsiloxane) (PDMS) or similar polymeric materials, which means that the microchannels in the devices are compliant \cite{LOVB97,XW98}.
Since at least the work by Gervais et al.\ \cite{GEGJ06}, it has been understood that the flow-induced deformation of microchannels can be significant. In turn, this deformation can alter the static and dynamic response of microfluidic devices, and it requires further mathematical analysis to accurately model and predict. This coupling is but one example of \emph{fluid--structure interactions} (FSIs), which occur at many scales across physics \cite{BGN14,DS16}. Over the past decade, a number of studies have begun to address FSIs in microfluidics, specifically by developing flow rate--pressure drop relationships for compliant microfluidic systems.

\begin{figure}[ht]
	\begin{center}
		\includegraphics[width=\columnwidth]{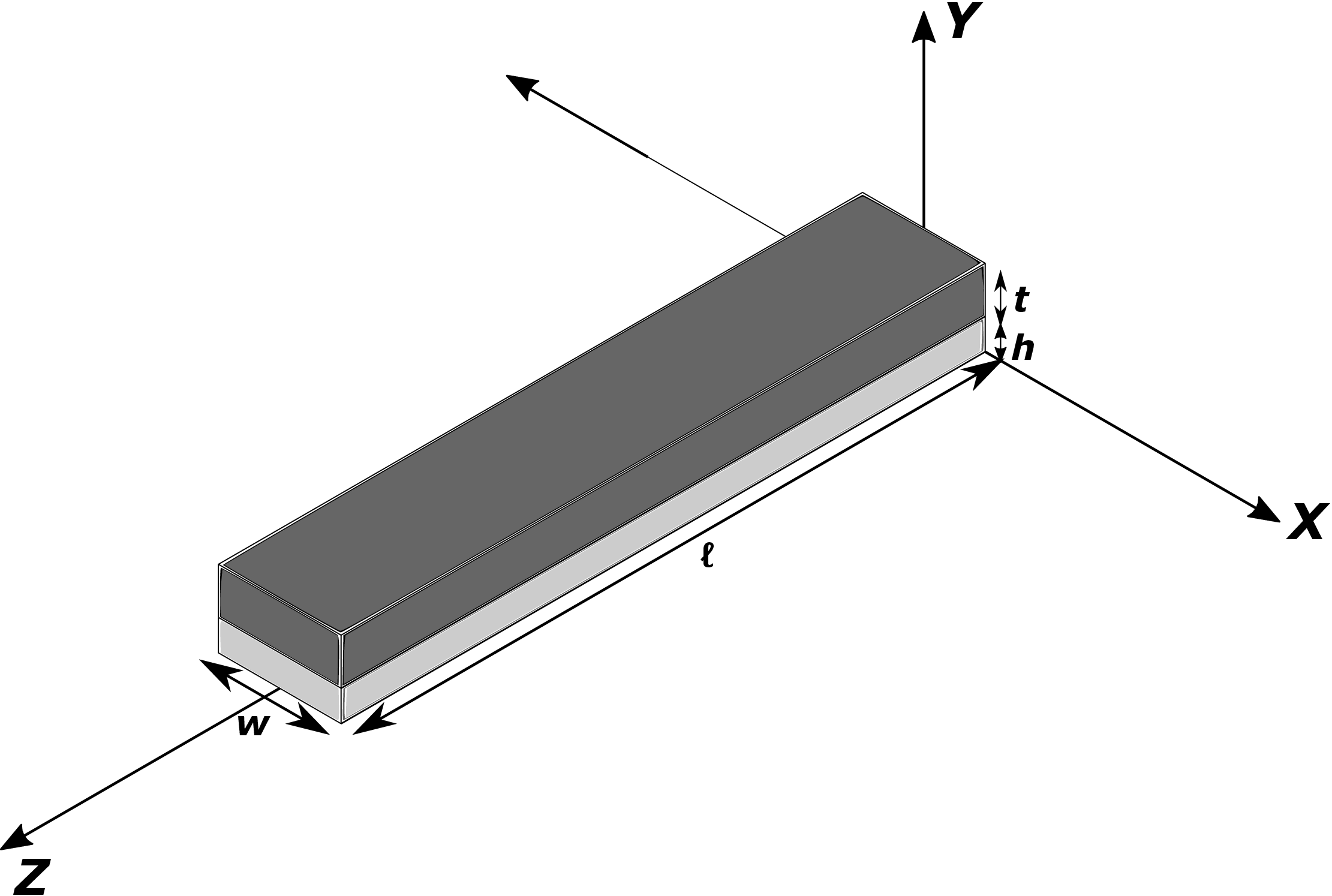}
	\end{center}
	\caption{Three-dimensional schematic of a typical microchannel geometry. The channel is long and thin, specifically its height $h$ ($=h_0$ for the undeformed configuration) is much smaller than its (fixed) width $w$, which is much smaller than its (fixed) length $\ell$. The flow-wise direction is $z$, with $x$ and $y$ being the cross-sectional (span-wise) coordinates. This geometry is also employed in the computational simulations below.}
	\label{fig:schematic}
\end{figure}

Through a scaling analysis, Gervais et al.~\cite{GEGJ06} characterized the behaviour of a rectangular microchannel, which was fabricated from PDMS, having a thick compliant top wall. This typical microchannel geometry (with the elastic top wall indicated by a darker shading) is shown in \fref{fig:schematic}. The volumetric flow rate $q$ through the microchannel was found to be a quartic function of the pressure $p(z)$ at any given downstream location $z$: 
\begin{equation}
\fl q = \frac{h_0^4 w}{48\alpha\mu(\ell-z)} \Bigg\{ \Bigg[1 + \frac{\alpha w}{Eh_0} p(z)\Bigg]^4  - 1\Bigg\},
\label{eq:q_p_gervais_dim}
\end{equation}
where $h_0$ is the undeformed channel height, $w$ is the fixed channel width, $\ell$ is the fixed channel length in the flow-wise direction, $E$ is the Young's modulus of PDMS, and $\mu$ is the fluid's dynamic viscosity. Here, $\alpha$ is a fitting parameter meant to capture some of the general features of the compliance of the system. The analysis in \cite{GEGJ06} posited a scaling relationship $u/w=\alpha p/E$, where $u$ is the cross-sectionally averaged microchannel top-wall displacement, in which $\alpha$ is not determined by the theory, and it must be calibrated against experiments in which \emph{both} $q$ and $p(z)$ are measured independently. The most pernicious feature of equation~\eref{eq:q_p_gervais_dim}, however, is that it is \emph{nonlinear} in $p(z)$, which is consistent with experimental measurements of flow in microchannels \cite{GEGJ06,SLHLUB09,HUZK09}.

A subsequent study by Hardy et al.~\cite{HUZK09} found a decrease of $\approx35\%$ in the pressure drop $\Delta p := p(0)-p(\ell)$ in compliant channels as compared to rigid-walled channels, which is in quantitative agreement with equation~\eref{eq:q_p_gervais_dim}. However, Hardy et al.~\cite{HUZK09} also reported that $\alpha$ depends on the thickness of the compliant wall. Thus, in general, $\alpha$ is a fitting parameter that depends on the geometric \emph{and} physical parameters of the system. It is of interest to determine this dependence {\it a priori}, without the need to fit a measured $q$ versus $\Delta p$ curve.

Next, Ozsun et al.~\cite{OYE13} discussed a modelling approach to deformable microchannels based on determining the maximum span-wise deformation using constitutive curves from hydrostatic bulge tests. Their approach showed good agreement with experiments but relied, much like the fitting parameter $\alpha$ from Gervais et al.~\cite{GEGJ06}, on fully characterizing the system experimentally before predictions could be made.

Meanwhile, Raj and Sen~\cite{RS16}, based on certain assumptions and correlations for elastic shells, proposed a relationship between $q$ and $p(z)$, which is no longer a polynomial: 
\begin{eqnarray}
\fl q = \frac{h_0^3 w\, p(z)}{12\mu(\ell-z)}\Bigg\{ 1 &+ \frac{9}{4}[f_\mathrm{p} p(z)]^{1/3} + \frac{9}{5}[f_\mathrm{p} p(z)]^{2/3} \nonumber\\
&+ \frac{1}{2}f_\mathrm{p} p(z) \Bigg\},
\label{eq:q_p_rs_dim}
\end{eqnarray}
where $f_\mathrm{p}$ is a ``compliance parameter'' introduced therein, given by 
\begin{equation}
f_\mathrm{p} = \frac{1}{72}\Bigg[\frac{w^4(1-\nu_\mathrm{s})}{h_0^3\, t E}\Bigg].\label{eq:fp}
\end{equation}
Here, $t$ is the thickness of the top wall (see \fref{fig:schematic}), and $\nu_\mathrm{s}$ is the Poisson ratio of the elastic material from which it is manufactured. Since this flow rate--pressure drop relation is based on \emph{stretching} of elastic shells, the scaling relationship between displacement and applied stress is of the form $u/w \propto (p/E)^{1/3}$, which leads to the non-integer powers of $p(z)$ in equation \eref{eq:fp}.

{More recently, Raj et al.~\cite{RDC17}, showed that the scaling assumption ($u/w=\alpha p/E$) used by Gervais et al.~\cite{GEGJ06} can be made more precise on the basis of correlations for the bending of plates \cite{TWK59}. Specifically, the ``thick-membrane approximation'' $u/w=\frac{1}{99}(w/t)^3(1-\nu_\mathrm{s}^2)(p/E)$ yields 
\begin{equation}
  \alpha = \frac{1}{99} (w/t)^3(1-\nu_\mathrm{s}^2)
\label{eq:alpha_raj}
\end{equation}
to be used in equation~\eref{eq:q_p_gervais_dim}. In other words, with the proper elasticity model at hand, $\alpha$ need not be a fitting parameter. Nevertheless, none of the studies summarized so far modelled the cross-sectional displacement profile; only cross-sectionally averaged displacements were used.}

In the most recent study by Christov et al.~\cite{CCSS17}, an analytical flow rate--pressure drop relationship was derived using perturbation theory under the lubrication approximation ($\ell\gg w\gg h$), starting from the equations of Stokes flow and the Kirchhoff--Love plate theory of elasticity. In this way, the leading-order (with the microchannel aspect ratio as the small parameter) velocity and displacement profiles were calculated. The resulting flow rate--pressure drop relation was found to be a quartic polynomial in $p(z)$ [like equation~\eref{eq:q_p_gervais_dim}] and to depend strongly on the bending rigidity $B$ of the compliant top wall:
\begin{eqnarray}
\fl q = \frac{h_0^3w\,p(z)}{12\mu(\ell-z)}\Bigg[ 1 &+ \frac{1}{480}\frac{w^4}{B h_0}p(z) \nonumber \\
&+ \frac{1}{362\,880}\left(\frac{w^4}{B h_0}\right)^2 p(z)^2 \nonumber\\ 
&+ \frac{1}{664\,215\,552}\left(\frac{w^4}{B h_0}\right)^3 p(z)^3 \Bigg],
\label{eq:q_p_icc_dim}
\end{eqnarray}
where $B = {Et^3}/[12(1-\nu_\mathrm{s}^2)]$.


{Here, we continue this line of research. Specifically, the novelty of the present work over the above-surveyed literature is as follows. We seek to determine ``how far''  expressions such as equations~\eref{eq:q_p_gervais_dim}, \eref{eq:q_p_rs_dim} and \eref{eq:q_p_icc_dim} can be ``pushed,'' and how to modify the theory self-consistently for different types of elastic deformation (e.g., thin versus thick plate-bending, stretching, etc.). To this end, in the present work, we perform some of the first detailed three-dimensional direct numerical simulations of static microchannel deformation. Furthermore, unlike previous theoretical work, here we benchmark the simulations and theory for the first time against two very different sets of experiments, namely those of Ozsun et al.~\cite{OYE13} and those of Raj and Sen \cite{RS16}. This approach not only allows for comparisons against actual experimental data, but it also supports our ``mission'' of testing the range of validity of the mathematical models in real-world situations. In doing so, we also develop a new extension of the theory of Christov et al.~\cite{CCSS17} to thick plates, bringing to light the dependence of the flow rate--pressure drop relation on the thickness-to-width ratio $t/w$.}

To this end, this paper is organized as follows. In section~\ref{sec:models}, we summarize the derivation of the fitting-parameter-free mathematical models that we employ in this work. These models are based on the approach of Christov et al.~\cite{CCSS17}, and we further extend the latter approach (in section~\ref{sec:bending_thick}) to thick-plate top walls. In section~\ref{sec:comp}, we discuss our computational modelling approach using ANSYS$^\textsuperscript{\textregistered}$ Workbench, including solver choices and parameters, boundary conditions, and grid refinement studies. Section~\ref{sec:properties} addresses the difficult question of how to find/estimate the numerical values of the material parameters of the PDMS walls used in microfluidic experiments. Then, section~\ref{sec:results} summarizes our main results. Specifically, we show good agreement between theoretical, experiment and simulation results on the flow rate--pressure drop relation (section~\ref{sec:results_qdp}); we compare the theoretical and computational results for the pressure distribution in the deformed channel (section~\ref{sec:results_pressure}); and, we present the first detailed comparison of the theoretical and computational predictions for the channel top wall's cross-sectional deformation profile (section~\ref{sec:results_deformation}). Finally, section~\ref{sec:conc} summarizes our main results and discusses future work and open questions. An appendix on elastic ``bulge tests'' is provided for completeness and to support our choices of material parameters in section~\ref{sec:properties}.

\section{Mathematical models}
\label{sec:models}

\subsection{The lubrication approximation for flow in a microchannel}
\label{sec:models_intro}

All the flow rate--pressure drop relation models discussed in \sref{sec:intro} are derived (explicitly or implicitly) on the basis of the lubrication approximation for flow in a microchannel. In particular, this means that the velocity field $\vec{v}=(v_x,v_y,v_z)$ of the fluid is primarily in the axial (i.e., $z$) direction and is given by (see, e.g., \cite[\S3.4.2]{B08}):
\begin{equation}
v_z(x,y,z) = -\frac{1}{2\mu}\frac{dp}{dz}\big[h(x,z)-y\big]y,
\label{eq:vz_lubri}
\end{equation}
where $h(x,z) = h_0 + u(x,z)$ is the deformed channel shape with $u(x,z)$ being the deformation. The pressure is solely a function of the flow-wise coordinate $z$. From equation \eref{eq:vz_lubri}, the volumetric flow rate follows, by definition:
\begin{eqnarray}
q &:= \int_{-w/2}^{+w/2}\int_0^{h(x,z)} v_z(x,y,z) \,dy\,dx \nonumber\\
  &= -\frac{1}{12\mu}\frac{dp}{dz} \int_{-w/2}^{+w/2} [h_0 + u(x,z)]^3 \, dx.
\label{eq:q_lubri}
\end{eqnarray}
When the latter integral can be evaluated, equation~\eref{eq:q_lubri} is an ordinary differential equation (ODE) for $p(z)$. This ODE is typically solved subject to $p(\ell)=0$, assuming the microchannel's exit is held at atmospheric pressure, which sets the pressure gage.

Next, an appropriate relationship must be found between the microchannel's deformation $u(x,z)$ and the hydrodynamic pressure $p(z)$. At this step, a variety of assumptions have been made \cite{GEGJ06,RS16,RDC17}. Typically, these involve relating the \emph{average} cross-sectional displacement (in a span-wise plane) to the pressure, which is independent of the cross-sectional coordinates to the leading-order in the lubrication approximation, via an engineering correlation for uniformly loaded plates or membranes (e.g., as summarized in \cite{S11}). However, because this approach does not provide the actual cross-sectional shape of the deflected top wall, the problem is not fully specified, hence the integral in equation \eref{eq:q_lubri} cannot be evaluated. Instead, at this point in the analysis, most of the current literature applies the scaling approach of Gervais et al.~\cite{GEGJ06} to complete the solution for the flow rate. In the next subsection, we discuss how to couple the lubrication flow discussed in this subsection to an appropriate elasticity problem. In doing so, we show how the problem of fluid--structure interactions in microchannels can be treated self-consistently within an asymptotic theory.

\subsection{Elastic response of the top wall and displacement profiles}

\subsubsection{Plate bending}
\label{sec:bending}

Assuming a maximum displacement  $u_\mathrm{max} \ll t \ll w$, Christov et al.~\cite{CCSS17} used perturbation methods to show that the flow-wise and transverse deformation of a thin linearly elastic plate decouple under the lubrication scaling $\ell\gg w\gg h_0$ discussed above. Then, at each fixed-$z$ cross-section, the problem of the top-wall deformation reduces to solving the Euler--Bernoulli beam equation subject to a uniform distributed load $p(z)$. The latter is easily solved \cite{CCSS17}, subject to clamped boundary conditions at $x=\pm w/2$, to arrive at
\begin{equation}
u(x,z) = \frac{u_{\mathrm{c},\mathrm{b}}}
{24} \left[\frac{p(z)}{\Delta p}\right] \left(\frac{x}{w}-\frac{1}{2}\right)^2 \left(\frac{x}{w}+\frac{1}{2}\right)^2,
\label{eq:u_bend_dim}
\end{equation}
where $u_{\mathrm{c},\mathrm{b}} = w^4\Delta p/B$ is the characteristic displacement due to bending. On substituting equation~\eref{eq:u_bend_dim} into \eref{eq:q_lubri} solving the ODE for $p(z)$ subject to $p(\ell)=0$, the flow rate--pressure drop relation \eref{eq:q_p_icc_dim} follows.

Introducing the dimensionless variables $U(X,Z) = u(x,z)/u_{\mathrm{c},\mathrm{b}}$, $X=x/w$, $Z=z/\ell$ and $P(Z)=p(z)/\Delta p$, equation \eref{eq:u_bend_dim} becomes 
\begin{equation}
\frac{U(X,Z)}{P(Z)} = \frac{1}{24}\left(X-{1}/{2}\right)^2\left(X+{1}/{2}\right)^2.
\label{eq:u_bend_dimless}
\end{equation}
In other words, the problem exhibits self-similarity \cite{B96}, and $U(X,Z)/P(Z)$ is a universal profile independent of the flow-wise direction.\footnote{In fact, it should be evident that this will be true for any type of top-wall elastic response as long as the flow-wise and span-wise deformations decouple asymptotically for $h_0\ll w\ll \ell$.} In \cite{CCSS17}, an additional dimensionless group ${\beta} := u_{\mathrm{c},\mathrm{b}}/h_0$ was introduced to characterize the compliance of the top wall; for $\beta\ll1$ the top wall is quite stiff, while for $\beta\gg1$ the top wall is quite flexible. Comparisons between the predicted self-similar profile [equation~\eref{eq:u_bend_dimless}] and full numerical simulations will be shown in \sref{sec:results_deformation}
 below.

\subsubsection{Thick-plate bending}
\label{sec:bending_thick}

If $t\simeq w$, i.e., the plate thickness is comparable to the width of the microchannel, then the Kirchhoff--Love plate-bending theory of the previous subsection does not apply. To account for the plate's nontrivial thickness, Mindlin's plate-bending theory \cite{M51} can be applied. The governing equations are expressed in terms of the deformation $u$, the rotation of the normal $\phi$, and the stress resultants $\mathcal{Q}$ and $\mathcal{M}$ (see, e.g., \cite[Chap.~13]{ZTZ13}). The equilibrium conditions are 
\begin{eqnarray}
\frac{\partial \mathcal{Q}}{\partial x} + p(z) &= 0,\label{eq:bend_mindlin1}\\
\frac{\partial \mathcal{M}}{\partial x} - \mathcal{Q} &= 0,\label{eq:bend_mindlin2}
\end{eqnarray}
and the corresponding constitutive relations are
\begin{eqnarray}
\mathcal{M} &= B\frac{\partial\phi}{\partial x},\label{eq:M_mindlin}\\
\mathcal{Q} &= \kappa G t \left(\frac{\partial u}{\partial x} + \phi\right),
\label{eq:Q_mindlin}
\end{eqnarray}
where $\kappa$ is a ``shear correction factor'' (see, e.g., \cite{GW01,H01,Z06} for a detailed discussion), and $G=E/[2(1+\nu_\mathrm{s})]$ is the elastic shear modulus. 

The plate's cross-section lies in the $(x,y)$ plane, and the in-plane displacements are given by $u$ and $y\phi$, respectively. Clamped conditions at $x=\pm w/2$ require that $u=\phi=0$ at $x=\pm w/2$. The solution to the system of ODEs \eref{eq:bend_mindlin1}--\eref{eq:Q_mindlin} subject to these boundary conditions is readily found to be
\begin{eqnarray}
u(x,z) = \frac{u_{\mathrm{c},\mathrm{b}}}{24} \left[\frac{p(z)}{\Delta p}\right] \left[\frac{1}{4} - \left(\frac{x}{w}\right)^2\right]\nonumber\\
\qquad \times \left\{ \frac{2(t/w)^2}{\kappa(1-\nu_\mathrm{s})}  + \left[\frac{1}{4} - \left(\frac{x}{w}\right)^2\right] \right\}, 
\label{eq:u_bend_thick}
\end{eqnarray}
where $u_{\mathrm{c},\mathrm{b}} = w^4\Delta p/B$ as in \Sref{sec:bending}. Clearly, as $t/w\to0^+$ the solution (\ref{eq:u_bend_thick}) reduces to that of (thin) plate bending (\ref{eq:u_bend_dim}), however, it has been pointed out that this convergence process is mathematically subtle and depends strongly on the type of deformation and the value of $\kappa$ \cite{Z06}.

For reference, from equation~\eref{eq:u_bend_thick}, it can be shown that
\begin{eqnarray}
u_\mathrm{max}(z) &:= \max_{-w/2\le x\le +w/2} u(x,z) = u(0,z) \nonumber\\
&=\frac{u_{\mathrm{c},\mathrm{b}}}{384} \left[\frac{p(z)}{\Delta p}\right]  \left[ 1+\frac{8(t/w)^2}{\kappa(1-\nu_\mathrm{s})} \right],
\label{eq:u_max_thick}
\end{eqnarray}
and
\begin{eqnarray}
\bar{u}(z) &:= \frac{1}{w}\int_{-w/2}^{+w/2} u(x,z)\, dx \nonumber\\
&= \frac{u_{\mathrm{c},\mathrm{b}}}{720} \left[\frac{p(z)}{\Delta p}\right]\left[1+\frac{10(t/w)^2}{\kappa(1-\nu_\mathrm{s})} \right].
\end{eqnarray}

Finally, on substituting equation~\eref{eq:u_bend_thick} into equation \eref{eq:q_lubri}, and solving the ODE for $p(z)$ subject to $p(\ell)=0$, the flow rate--pressure drop relation for the case of a thick-plate top wall follows:
\begin{eqnarray}
\fl q &= \frac{h_0^3w\,p(z)}{12\mu(\ell-z)}\Bigg\{ 1 + \left[\frac{1}{480}+f_1(t/w)\right]\frac{w^4}{B h_0}p(z) \nonumber \\
&\phantom{=} + \left[\frac{1}{362\,880}+f_2(t/w)\right]\left(\frac{w^4}{B h_0}\right)^2 p(z)^2 \nonumber\\ 
&\phantom{=} + \left[\frac{1}{664\,215\,552}+f_3(t/w)\right]\left(\frac{w^4}{B h_0}\right)^3 p(z)^3 \Bigg\} ,
\label{eq:q_p_thick_dim}
\end{eqnarray}
where 
\begin{eqnarray}
f_1 &:=\frac{(t/w)^2}{48\kappa(1-\nu_\mathrm{s})}, \label{eq:f1}\\
f_2 &:= \frac{(t/w)^2}{20\,160\kappa(1-\nu_\mathrm{s})} + \frac{(t/w)^4}{4320[\kappa(1-\nu_\mathrm{s})]^2}, \label{eq:f2}\\
f_3 &:= \frac{(t/w)^2}{25\,546\,752\kappa(1-\nu_\mathrm{s})} +\frac{(t/w)^4}{2\,903\,040[\kappa(1-\nu_\mathrm{s})]^2} \nonumber \\
 &\phantom{:=} + \frac{(t/w)^6}{967\,680[\kappa(1-\nu_\mathrm{s})]^3}. \label{eq:f3}
\end{eqnarray}
Note that this result is fundamentally different from the thick-plate flow rate--pressure drop relation derived in \cite{RDC17} based on only a correlation between maximum displacement and pressure, following the approach of Gervais et al.~\cite{GEGJ06}. Furthermore, notice that the corrections to equation~\eref{eq:q_p_icc_dim}, embodied by the functions $f_1$, $f_2$ and $f_3$ present in equation~\eref{eq:q_p_thick_dim}, can be as large as the numerical factors already in the brackets. For example, for a plate with $t/w \simeq 0.32$ (similar to the ``OZ5'' case to be introduced below) with $\nu_\mathrm{s}=0.5$ and $\kappa=1$, $f_1=1/240$ whereas the numerical factor being corrected is $1/480$.

Using the dimensionless variables introduced before equation~\eref{eq:u_bend_dimless}, equation~\eref{eq:u_bend_thick} becomes
\begin{eqnarray}
\frac{U(X,Z)}{P(Z)} = \frac{1}{24} \left(\frac{1}{4} - X^2\right) \nonumber\\
\qquad\times\left[ \frac{2(t/w)^2}{\kappa(1-\nu_\mathrm{s})}  + \left(\frac{1}{4} - X^2\right) \right].
\label{eq:u_bend_thick_dimless}
\end{eqnarray}

\subsubsection{Stretching: Approximations}

When the maximum displacement $u_\mathrm{max}\gg t$, the elastic response is predominantly that of membrane stretching rather than plate bending. Although we do not have strong  evidence of such an elastic response in deformable microchannels (under typical loading conditions), Raj and Sen \cite{RS16} nevertheless assumed this regime in their model.

To account for stretching, the F\"oppl--von K\'{a}rm\'{a}n equations must be employed \cite{TWK59,HKO09}. However, these equations are inherently nonlinear and, unlike the plate-bending equations considered in sections~\ref{sec:bending} and \ref{sec:bending_thick}, they do not possess applicable exact solutions. Therefore, the displacement profile $u(x,z)$ cannot be computed exactly, and it is not expected that a ``nice'' power-law dependence on the pressure, which allowed for the self-similar rescaling in equations~\eref{eq:u_bend_dimless} and \eref{eq:u_bend_thick_dimless}, can be obtained. 

Raj and Sen \cite{RS16} used an approximation based on the discussion of \emph{thin circular membranes} in \cite{S11}, specifically, they posited that the maximum displacement at $x=0$ is given by
\begin{equation}
u_\mathrm{max}(z) = u(0,z) \approx \left(\frac{3}{64}\right)^{1/3} u_{\mathrm{c},\mathrm{s}} \left[ \frac{p(z)}{\Delta p} \right]^{1/3},
\label{eq:u_max_rs}
\end{equation}
where $u_{\mathrm{c},\mathrm{s}} = [w^4\Delta p (1-\nu_\mathrm{s})/(t E)]^{1/3}$. Equation~\eref{eq:u_max_rs}, when used within the approach of Gervais et al.~\cite{GEGJ06}, yields the flow rate--pressure drop relation given by equation~\eref{eq:q_p_rs_dim}. However, we also note that based on the assumption of a parabolic displacement profile in \cite{RS16}, and using equation~\eref{eq:u_max_rs}, we would have 
\begin{equation}
u(x,z) \approx \left(\frac{3}{64}\right)^{1/3} u_{\mathrm{c},\mathrm{s}} \left[ \frac{p(z)}{\Delta p} \right]^{1/3} \left[1 - 4 \left(\frac{x}{w}\right)^2\right].
\label{eq:u_stretch_rs}
\end{equation}
The latter expression could then be substituted into equation~\eref{eq:q_lubri} to calculate a flow rate--pressure drop relation, \emph{different} from equation~\eref{eq:q_p_rs_dim}, based on our approach outlined in section~\ref{sec:models_intro}.

\subsection{Channels that are not necessarily shallow}

Finally, for completeness, we note that Cheung et al.~\cite{CTS12} extended the approach of Gervais et al.~\cite{GEGJ06} to account for channels with aspect ratios $h_0/w$ close to $1$ (i.e., square cross-sections) by introducing an additional term in the $q$--$p(z)$ relationship:
\begin{eqnarray}
\fl q &= \frac{h_0^4 w}{48\alpha\mu(\ell-z)} \Bigg\{ \Bigg[ \Bigg(1 + \frac{\alpha w}{Eh_0}p(z)\Bigg)^4  - 1\Bigg] \nonumber\\
 &\phantom{=} - 0.504\,\frac{h_0}{w}\Bigg[ \Bigg(1 + \frac{\alpha w}{Eh_0}p(z)\Bigg)^5  - 1\Bigg]\Bigg\}.
\label{eq:q_p_cheung_dim}
\end{eqnarray}
This approach was also used in \cite{RS16,RDC17}. 

Christov et al.~\cite{CCSS17}, however, argued that this is not consistent with the perturbation approach and may overestimate the effect of lateral side walls. The asymptotically consistent version of equation~\eref{eq:q_p_cheung_dim} simply involves subtracting off the value $\varkappa_0 \cdot (h_0/w)$, where $\varkappa_0\approx 0.630$, from the quantity in the brackets in, e.g., equation~\eref{eq:q_p_icc_dim} \cite{CCSS17}. That said, the first correction in $h_0/w$ is negligible for all three data sets considered herein, hence we do not consider these modified flow rate--pressure drop relations.


\section{Computational model}
\label{sec:comp}

Fluid--structure interactions (FSIs) are a topic at the forefront of computational mechanics, with many studies focusing on the efficient numerical solution of large-scale FSI problems \cite{HWL12}. Nevertheless, only a few previous studies have performed three-dimensional (3D) simulations of the types of FSIs relevant to microfluidic devices, specifically soft microchannels. Gervais et al.~\cite{GEGJ06} performed some simulations to compare to their experiments. More recently, Chakraborty et al.~\cite{CPFY12} conducted a detailed numerical study, including the effect of viscoelasticity of the soft solid wall. In the present work, we conduct 3D, two-way coupled FSI simulations in order to benchmark the mathematical models above and to obtain reliable data on the cross-sectional displacement profile of the microchannel's top wall.

\subsection{Microchannel geometry}
\label{sec:comp_geo}
        
To this end, we consider a rectangular channel with dimensions $h_0 \times w \times \ell$. A 3D sketch is shown in \fref{fig:schematic}. The top wall (dark shaded region) of the channel has a thickness $t$ and is made of a compliant material (PDMS in our case). Microchannel S4 from \cite{OYE13} (hereafter labelled as ``OZ4'') has been considered as the benchmark case as its experimental conditions were found to be within the scope of the assumptions of the asymptotic modelling approach from \cite{CCSS17} discussed above. Two additional microchannel geometries, from \cite{OYE13} and \cite{RS16} labelled hereafter as ``OZ5'' and ``RS1,'' respectively, were also selected to test the extent to which the bending-dominated mathematical models can be ``pushed.'' Table \ref{tb:dims} lists the undeformed channel dimensions for each case.

\begin{table}
\caption{\label{tb:dims}Dimensions of the microchannels for each of the three chosen benchmark experiments.}
\footnotesize\rm
\begin{tabular*}{\columnwidth}{@{}l*{15}{@{\extracolsep{0pt plus12pt}}l}}
\br
Case & $\ell$ (mm) & $w$ ($\mu$m) & $h_0$ ($\mu$m) & $t$ ($\mu$m)\\
\mr
OZ4&15.5&1700&244&200\\
OZ5&15.5&1700&155&605\\
RS1&36&500&83&55\\
\br
\end{tabular*}
\end{table}

\subsection{Fluid--structure interaction methodology} 
The approach chosen to model any FSI problem strongly depends on the degree/type of coupling being enforced between the fluid and the solid fields \cite{HWL12}. A monolithic approach is one in which the discretized governing equations of the fluid and the solid fields are assembled into a single matrix equation. While this provides for tight coupling between the two fields, enforcing the governing equations together faithfully, solving this large system of algebraic equations efficiently is challenging, and good performance for 3D problems can only be achieved through the development of problem-specific pre-conditioners \cite{HHB08}. 

On the other hand, one-way coupling provides a numerically inexpensive approach by assuming that only one of the two fields depends on the other. This approach is only suitable for situations where either the fluid or solid physics dominates the problem. In the present context, the flow within the microchannel and the deformation of the top wall are altered significantly by each other, hence they are intrinsically coupled. 

\begin{figure}[ht]
	\begin{center}
		\includegraphics[width=0.9\columnwidth]{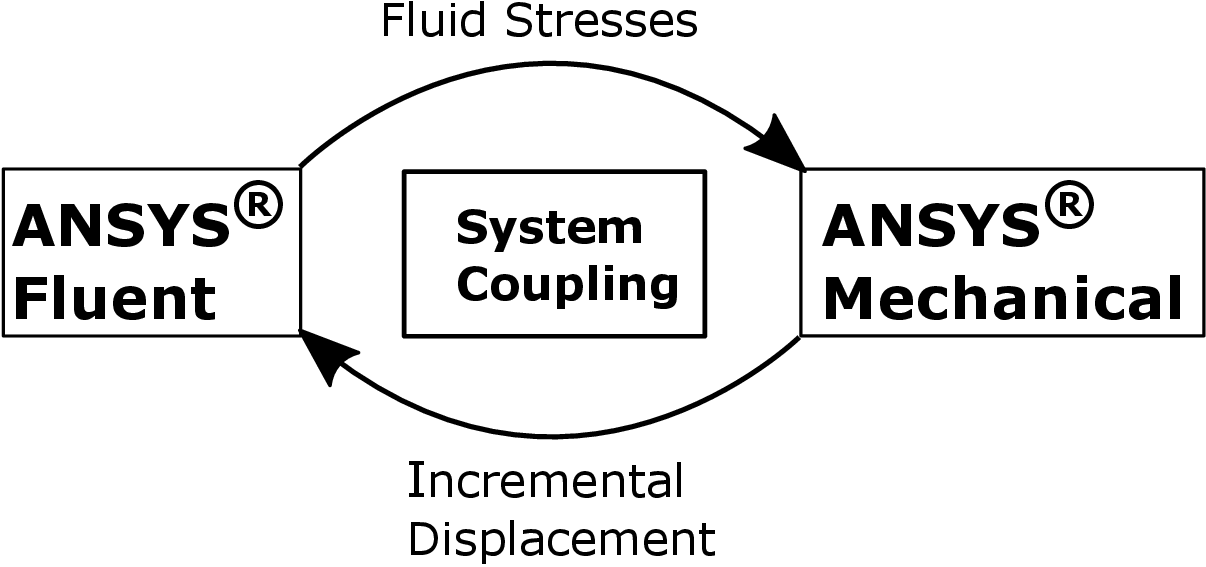}
	\end{center}
	\caption{Schematic of the two-way coupling procedure for FSI in ANSYS$^\textsuperscript{\textregistered}$ Workbench.}
	\label{fig:ansys_fsi}
\end{figure}

To capture the physics of flow-induced microchannel deformation, we thus chose a two-way coupling strategy, which proves to be more computationally efficient than a monolithic approach while maintaining  fidelity to the problem physics (unlike one-way coupling). To this end, the microchannel is partitioned into a fluid domain and a solid domain, these being the fluid inside the channel and the compliant top wall, respectively. The commercial computer-aided engineering (CAE) platform ANSYS$^\textsuperscript{\textregistered}$ Workbench was used for the simulations. Specifically, ANSYS$^\textsuperscript{\textregistered}$ Fluent and Mechanical were chosen to solve for the fluid and solid fields, respectively. The two fields were coupled using the built-in ``System Coupling'' module. A schematic of the work-flow of the simulation and data transfers is shown in \fref{fig:ansys_fsi}.

\subsection{Solver details and parameters}
In this subsection, we summarize the details pertaining to the fluid and and solid mechanics solvers. Owing to the fact that the solvers used are commercial, most details of their implementation are propriety to ANSYS$^\textsuperscript{\textregistered}$. However, we have provided as much information as possible to ensure that our simulations can be reproduced.

\subsubsection{ANSYS$^\textsuperscript{\textregistered}$ Fluent}
A pressure-based $\vec{v}$--$p$ coupled solver is used to solve the steady-state incompressible Navier--Stokes equations for a Newtonian fluid:
\begin{eqnarray}
\nabla\cdot\vec{v} = 0, \label{eq:NS_cont}\\
\nabla\cdot(\vec{v}\otimes\vec{v}) &= -\frac{1}{\rho}\nabla p + \nu \nabla^2 \vec{v}, \label{eq:NS_mom}
\end{eqnarray}
where $\vec{v}(x,y,z)$ is the fluid's velocity field, $p(x,y,z)$ is the hydrodynamic pressure, $\rho$ is the fluid's (constant) density, $\nu=\mu/\rho$ is the kinematic viscosity, and $\vec{v}\otimes\vec{v}$ is the dyadic product of two vectors. For improved stability and convergence properties, equations \eref{eq:NS_cont} and \eref{eq:NS_mom} are discretized in a ``pseudo-transient'' formulation in which implicit under-relaxation is used to obtain the steady state.

A second-order upwind scheme was chosen for the spatial discretization of the momentum fluxes. For pressure interpolation, a second-order scheme was used. Gradient evaluations in the discretized equations were carried out using the Green--Gauss node-based approach. Uniform convergence criteria of $10^{-6}$ were used for the scaled residual values. In order to handle the deformation of the fluid mesh, dynamic meshing was used. The interior volume mesh was designated as a deformable dynamic zone and spring-based smoothing and remeshing methods were employed to update the mesh. Additionally, the top wall was designated as a ``system coupling'' type zone, whose behaviour was dictated by the System Coupling module. Additional information regarding the solver, the dynamic mesh update methods and each discretization scheme can be found in the ANSYS$^\textsuperscript{\textregistered}$ Fluent Theory Guide \cite{ANSYS}.

\subsubsection{ANSYS$^\textsuperscript{\textregistered}$ Structural}

Using the standard finite element approach, the discretized static force-balance equations for each element can be assembled into a matrix equation:
\begin{equation}
\{F_\mathrm{a}\} = \mathbf{K} \{u\},
\label{eq:Solid_eqn_fem}
\end{equation}
where {$F_\mathrm{a}$} is the applied load vector, $\mathbf{K}$ is the global stiffness matrix and $\{u\}$ is the unknown vector containing the values of the prescribed degrees of freedom for each node. The constitutive behaviour is assumed to be that of a linear, isotropic elastic solid, which in the finite element notation takes the form
\begin{equation}
\{\sigma\} = \mathbf{D}\{\varepsilon\},
\label{eq:Solid_eqn}
\end{equation}
where $\{\sigma\}$ is the vector containing the stress components, $\mathbf{D}$ is the elasticity matrix, and $\{\varepsilon\}$ is the corresponding vector of strains.

A large-deformation formulation is employed, making the strain $\{\varepsilon\}$ a nonlinear function of the displacement $\{u\}$. Thus, the stiffness matrix $\mathbf{K}$ is a function of $\{u\}$. Hence, equation~\eref{eq:Solid_eqn_fem} is a nonlinear algebraic equation, and it is solved using a line-search-based Newton--Raphson method with the default convergence criteria (a relative error of $0.5\%$ for the force and moment balances). A sparse direct solver was used to solve the matrix equation at each Newton--Raphson iteration.

\subsection{Grid arrangement}

For the fluid domain, a finite volume Cartesian mesh was constructed. For the solid domain, a Cartesian finite element mesh was generated using a 20-node 3D element. In order to obtain computationally inexpensive meshes for OZ4 and OZ5 with the desired spatial resolution, the node density in both the fluid and solid meshes was selectively refined towards the fluid-solid interface. 

\subsection{Boundary conditions}

Boundary conditions were chosen to match the given experimental conditions as closely as possible. The side walls and the base of the channel were assumed to be perfectly rigid. The specified flow rates were imposed at the channel inlet, and the channel outlet was chosen as the pressure datum (zero gauge pressure). Ozsun et al.~\cite{OYE13} mention the use of a rigid microfluidic connection between the syringe pump and the test micro-channel section. In order to ascertain if the flow at the channel inlet is fully developed, the entrance length $L_\mathrm{d}$ for the highest flow rate was calculated using the standard empirical correlation \cite[p.~421]{ID96} for the development length of a laminar flow in a pipe with non-circular cross-section:
\begin{equation}
L_\mathrm{d} \approx 0.05D_\mathrm{i}Re. 
\label{eq:entrance_length}
\end{equation}
Here, $Re$ is the Reynolds number, and $D_\mathrm{i}$ is the hydraulic diameter of the pipe, both of which are calculated at the channel inlet as
\begin{eqnarray}
	Re &= \frac{4\rho Q}{\mu P_\mathrm{i}}, \\
	D_\mathrm{i} &= \frac{4A_\mathrm{i}}{P_\mathrm{i}},
\end{eqnarray}
where $A_\mathrm{i}$ is the inlet cross-sectional area, and $P_\mathrm{i}$ is the inlet cross-sectional perimeter. 

From equation~\eref{eq:entrance_length}, the entrance length  is estimated to be $L_\mathrm{d}\approx1$ mm for both the OZ4 and OZ5 data sets. It is reasonable to assume that the rigid connections used in \cite{OYE13} are longer than this $L_\mathrm{d}$, hence a fully-developed velocity profile was imposed for OZ4 and OZ5 by writing a user-defined function (UDF) in ANSYS$^\textsuperscript{\textregistered}$ Fluent. The first four terms of the exact series solution for flow in a rectangular ($h_0\times w$) channel given in \cite[\S3.4.6]{B08} were used to define the inlet velocity profile. Similarly, the RS1 simulations were performed at constant flow rate by imposing both a uniform inlet velocity (consistent with the experimental setup in \cite{RS16}) as well as with a fully-developed inlet velocity profile. In all cases, the compliant top wall was clamped along its edges. Clamping was achieved by imposing zero displacement along the side walls and inlet/outlet planes. The shared surface between the solid and the fluid domains was designated as a ``fluid-solid'' interface, which served as the location for stress--displacement data transfer between the two solvers.

\subsection{Grid convergence analysis and computational time}

In order to ensure that our simulation results were independent of the grid resolution, a grid convergence study was carried out for all three microchannel cases (OZ4, OZ5 and RS1). For a given flow rate and Young's modulus of PDMS, three different grid sizes were chosen for the solid and fluid domain each, making a total of nine grid combinations per microchannel system. Table \ref{tb:GS} lists the details for each grid for  OZ4. The remaining microchannels have been discretized using a similar meshing scheme. 

\begin{table}[ht]
\caption{Mesh parameters for the grid convergence study performed on the OZ4 data set. Three different solid grids and three different fluid grids, for a total of nine different combinations, were used. The node spacing (cell size) is denoted by $\mathfrak{h}$ and is given in $\mu$m.}
\footnotesize\rm
\begin{tabular}{@{\extracolsep{-6.5pt}}*{8}{l}}
\br 
&& \multicolumn{3}{c}{Fluid} & \multicolumn{3}{c}{Solid}\\
\ns\ns && \crule{3} & \crule{3}\\
axis && grid 1 & grid 2 & grid 3 & grid 1 & grid 2 & grid 3\\
\mr
\multirow{3}{*}{$x$} & min.\ $\mathfrak{h}$ & {10} & {14.9} & {20} & {25} & {30.4} & {39.5}\\
& max.\ $\mathfrak{h}$ & {10} & {14.9} & {20} & {25} & {30.4} & {39.5}\\
& \# divisions & {170} & {114} & {85} & {68} & {56} & {43}\\
\mr
\multirow{3}{*}{$y$} & min.\ $\mathfrak{h}$ & {4.3} & {5} & {7.6} & {4.4} & {10} & {12.4}\\
& max.\ $\mathfrak{h}$ & {45.6} & {50} & {58} & {67.5} & {75} & {85}\\
& \# divisions & {13} & {12} & {10} & {8} & {6} & {5}\\
\mr
\multirow{3}{*}{$z$} & min.\ $\mathfrak{h}$ & {15} & {20} & {20} & {25} & {30} & {40}\\
& max.\ $\mathfrak{h}$ & {15} & {20} & {20} & {25} & {30} & {40}\\
& \# divisions & {1034} & {775} & {775} & {620} & {517} & {388}\\
\br
\end{tabular}
\label{tb:GS}
\end{table}

\begin{figure}[t]
\begin{center}
	\includegraphics[width=0.9\columnwidth]{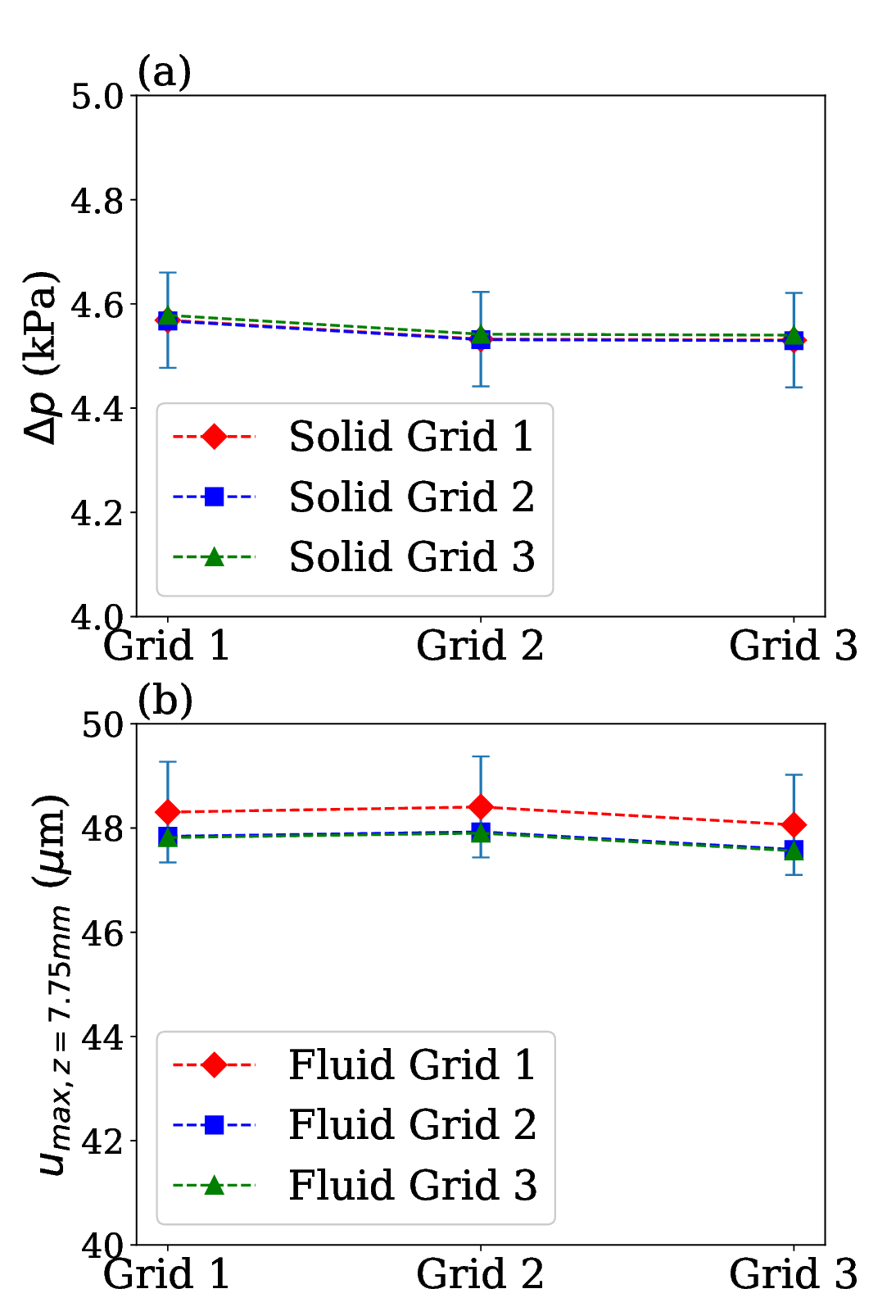}
\end{center}
\caption{Grid refinement study results for the OZ4 data set. (a) The total pressure drop $\Delta p$ across the microchannel as a function of the average fluid grid's spacing and for three different solid grids. (b) The maximum top-wall displacement at the channel's stream-wise half-length ($z=7.75$ mm) as a function of the average solid grid spacing and for three different fluid grids. The error bars represent a 2\% variation from the value of the finest grid spacing.}
\label{fig:GS_fluid}
\end{figure}

Figure \ref{fig:GS_fluid} shows the variation of (a) the full pressure drop and (b) the maximum deformation at the channel's half-length with each grid configuration at a flow rate of 50 mL/min for OZ4. The fluid properties are those of water, as discussed in \sref{sec:properties}, while $E=1.8$ MPa and $\nu_\mathrm{s}=0.499$ were used for the material properties of PDMS. The error bars represent a 2\% variation from the numerical value at the finest grid spacing (i.e., grid 1). As seen from the figure, no significant change is observed with a refinement of the grid. Hence, we can be confident that the results obtained fall within the range in which the numerical solution can be considered ``converged,'' and any further refinement would only lead to a significant increase in the computational cost/run time without any appreciable gains in accuracy.

{The total CPU computational time for each simulations depends strongly on the particular computational mesh and the number of cores engaged, which can vary based on availability. For the case of OZ4, the simulation run time for the highest flow rate with 10 cores was approximately 1 hour. For OZ5 and RS1, the corresponding run times were approximately 2 hours and approximately 37 minutes, respectively.}

\section{Material properties}
\label{sec:properties}

In order to obtain accurate simulation results, the material properties of both the fluid and solid have to be specified precisely. For all three cases, the fluid was taken to be water at $24^{\circ}$C, which is (to a very good approximation) an incompressible fluid with constant density $\rho = 997.3$ kg/m$^{3}$ and constant viscosity $\mu = 9.14 \times 10^{-4}$ Pa$\cdot$s. 

The material properties of PDMS remain the topic of active research \cite{ALA99,MFR05,LSC09,LSSBC09,JMTT14}, with some studies even suggesting a nonlinear mechanical response \cite{KKJ11}. As a result, determining the linear elastic properties (i.e., Young's modulus and Poisson ratio) of the PDMS top wall for our three data sets was challenging. In particular, several studies \cite{JMTT14,LSC09,LSSBC09} have shown that the Young's modulus $E$ and Poisson ratio $\nu_\mathrm{s}$ of PDMS strongly depend on the method of fabrication and parameters such as the curing temperature, the specimen thickness and the mixing ratio of the components of the polymeric solution. This dependence, in turn, can result in a typical variation of $E=1$ to 3 MPa and $\nu_\mathrm{s}=0.4$ to $0.5$. For the present analysis, the variation in the Poisson ratio has been neglected and PDMS has been modelled as a nearly incompressible solid ($\nu_\mathrm{s}=0.499$). 

Raj and Sen~\cite{RS16} have used a value of $E=1.362$ MPa, based on experimental results from \cite{LSSBC09}, for their theoretical calculations. However, knowing the sensitivity of the Young's modulus to curing conditions, we made an attempt to find the particular curing conditions under which Liu et al.~\cite{LSSBC09} have reported their measurements. From the corresponding thesis \cite{L08}, we found the specific curing temperature and duration to be 80$^{\circ}$C and $1.5$ hours, respectively. Taking into account the curing conditions in \cite{RS16} (65$^{\circ}$C for $3.5$ hours) and factoring-in the linear relationship between the Young's modulus and the curing temperature given in \cite{JMTT14}, we estimated a value of $E=1.17$ MPa for RS1. For OZ4 and OZ5, a value of $1.6$ MPa (see the appendix) was chosen. A variation of $\pm 0.2$ MPa ($\approx 12.5-17\%$) has been considered to account for any variability in $E$ due to unaccounted factors in the experimental conditions and manufacturing techniques.

Tables \ref{tb:params1}--\ref{tb:params3} list the values that we have estimated for the material constants ($E$ and $\nu_\mathrm{s}$). The flow rate $q$, the dimensionless FSI parameter $\beta := u_\mathrm{c,b}/h_0 \equiv w^4\Delta p/(Bh_0)$ (with $\tilde{\beta}:=\beta/24$ and $\Delta p$ estimated by the viscous pressure scale $\mu\ell q/(h_0^3w)$, see also the relevant discussion in section~\ref{sec:results_pressure} below) introduced in \cite{CCSS17}, the compliance parameter $f_\mathrm{p}$ from \cite{RS16} given by equation~\eref{eq:fp}, the best-fit value of the parameter $\alpha$ from \cite{GEGJ06} to be used in equation~\eref{eq:q_p_gervais_dim}, and (for comparison purposes) the predicted value of $\alpha$ by equation~\eref{eq:alpha_raj} based on \cite{RDC17}, are also given in these tables for completeness and future reference.

\begin{table}
\caption{\label{tb:params1}Material constants and flow parameters for the OZ4 data set and simulations.}
\footnotesize\rm
\begin{tabular}{@{}*{6}{l}}
\mr
$E$ (MPa) & \multicolumn{5}{l}{{1.6 $\pm$ 0.2}}\\
$\nu_\mathrm{s}$ & \multicolumn{5}{l}{0.499}\\ 
$\alpha$ [fit] & \multicolumn{5}{l}{1.7285}\\
$\alpha$ [eq.~\eref{eq:alpha_raj}] & \multicolumn{5}{l}{4.6525}\\
\mr
$q$ (mL/min) & {10} & {20} & {30} & {40} & {50}\\
$\beta$ & 2.301 & 4.602 & 6.903 & 9.205 & 11.51\\
$\tilde\beta$ & 0.09588 & 0.1918 & 0.2876 & 0.3835 & 0.4794\\
\br
\end{tabular}
\end{table}

\begin{table}
\caption{\label{tb:params2}Material constants and flow parameters for the OZ5 data set and simulations.}
\footnotesize\rm
\begin{tabular}{@{}*{6}{l}}
\mr
$E$ (MPa) & \multicolumn{5}{l}{{1.6 $\pm$ 0.2}}\\
$\nu_\mathrm{s}$ & \multicolumn{5}{l}{0.499}\\ 
$\alpha$ [fit] & \multicolumn{5}{l}{$-0.9333$}\\
$\alpha$ [eq.~\eref{eq:alpha_raj}] & \multicolumn{5}{l}{0.1681}\\
\mr
$q$ (mL/min) & {10} & {20} & {30} & {40} & {50}\\
$\beta$ & 0.511 & 1.022 & 1.534 & 2.045 & 2.556\\
$\tilde\beta$ & 0.0213 & 0.0426 & 0.0639 & 0.0852 & 0.1065\\
\br
\end{tabular}
\end{table}

\begin{table}
\caption{\label{tb:params3}Material constants and flow parameters for the RS1 data set and simulations.}
\footnotesize\rm
\begin{tabular}{@{\extracolsep{-2pt}}*{7}{l}}
\mr
$E$ (MPa) & \multicolumn{6}{l}{{1.17 $\pm$ 0.2}}\\
$\nu_\mathrm{s}$ & \multicolumn{6}{l}{0.499}\\ 
$\alpha$ [fit] & \multicolumn{6}{l}{48.5} \\
$\alpha$ [eq.~\eref{eq:alpha_raj}] & \multicolumn{6}{l}{5.6918}\\
$f_\mathrm{p}$ (Pa$^{-1}$)& \multicolumn{6}{l}{{$1.248\times10^{-5}$}}\\ 
\mr
$q$ (mL/min) & {0.1} & {0.3} & {0.5} & {0.7} & {1.0} & {1.2} \\
$\beta$ & 6.687 & 50.06 & 33.44 & 46.81 & 66.87 & 80.25\\
$\tilde\beta$ & 0.2783 & 0.8348 & 1.391 & 1.948 & 2.783 & 3.339\\
\br
\end{tabular}
\end{table}


\begin{figure*}[h!]
\begin{center}
	\includegraphics[width=0.9\textwidth]{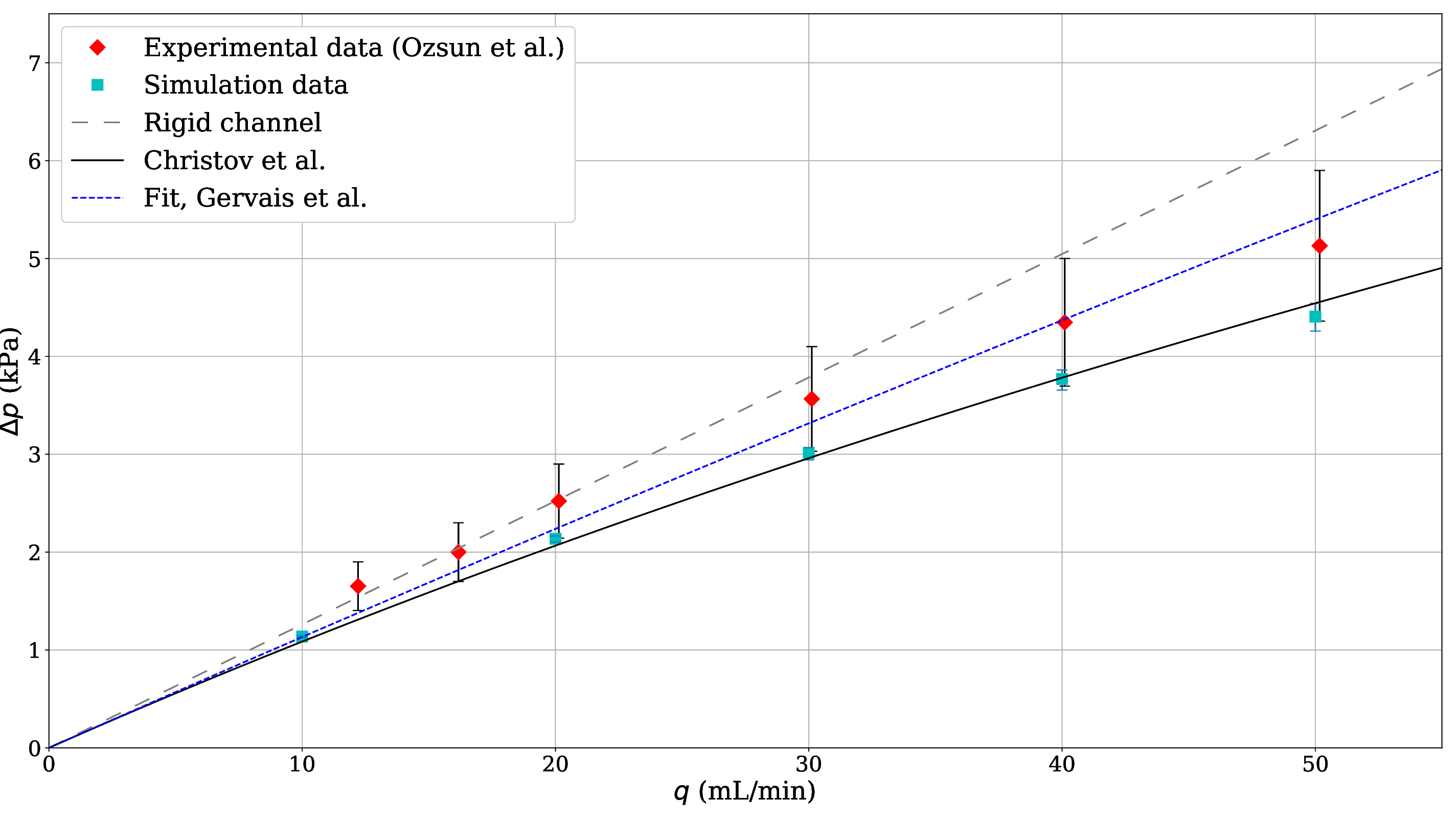}
\end{center}
\caption{$q$ versus $\Delta p$ for the data set OZ4. Error bars on the simulation data represent a $\pm$0.2 MPa variation in the Young's modulus of PDMS. Error bars on the experimental data represent a 15\% error margin, meant to emulate the statement in \cite{OYE13} that error bars are on the order of (or smaller) than the symbols used in the plots therein. A nonlinear least squares fit gives $\alpha=1.7285$ (see table~\ref{tb:params1}) to be used in equation~\eref{eq:q_p_gervais_dim}. The theoretical prediction from Christov et al.~\cite{CCSS17} is given by equation~\eref{eq:q_p_icc_dim}.}
\label{fig:q-dp_OZ4}
\end{figure*}

\section{Results}
\label{sec:results}

\subsection{Flow rate--pressure drop curves}
\label{sec:results_qdp}

\begin{figure*}
\begin{center}
	\includegraphics[width=0.9\textwidth]{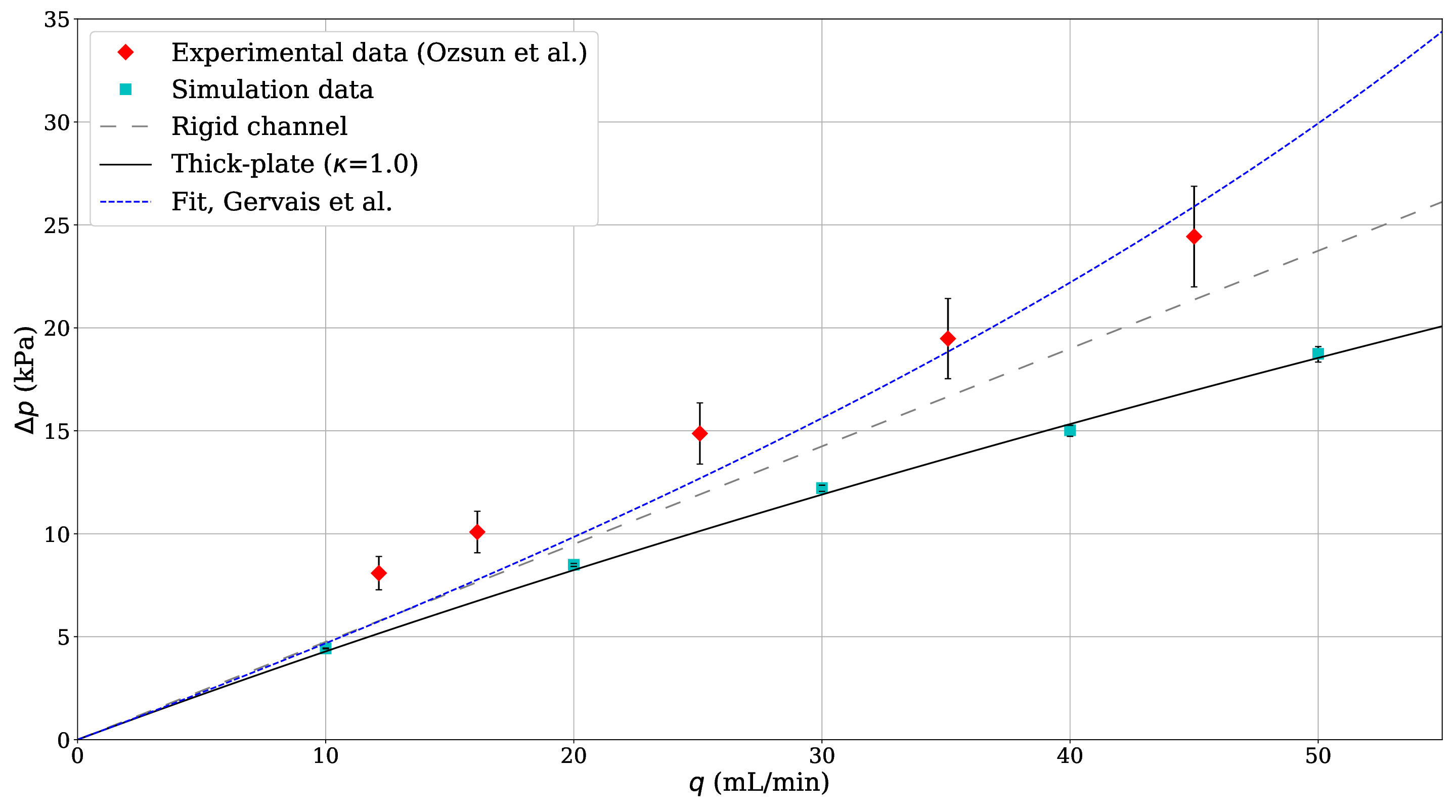}
\end{center}
\caption{$q$ versus $\Delta p$ for data set OZ5. Error bars on the simulation data represent a $\pm$0.2 MPa variation in the Young's modulus of PDMS. Error bars on the experimental data represent a 10\% error margin, meant to emulate the statement in \cite{OYE13} that error bars are on the order of (or smaller) than the symbols used in the plots therein. A nonlinear least squares fit gives $\alpha=-0.933$ (see table~\ref{tb:params2}) to be used in equation~\eref{eq:q_p_gervais_dim}. The thick-plate theoretical prediction is given by equation~\eref{eq:q_p_thick_dim}.}
\label{fig:q-dp_OZ5}
\end{figure*}

\begin{figure*}
\begin{center}
	\includegraphics[width=0.9\textwidth]{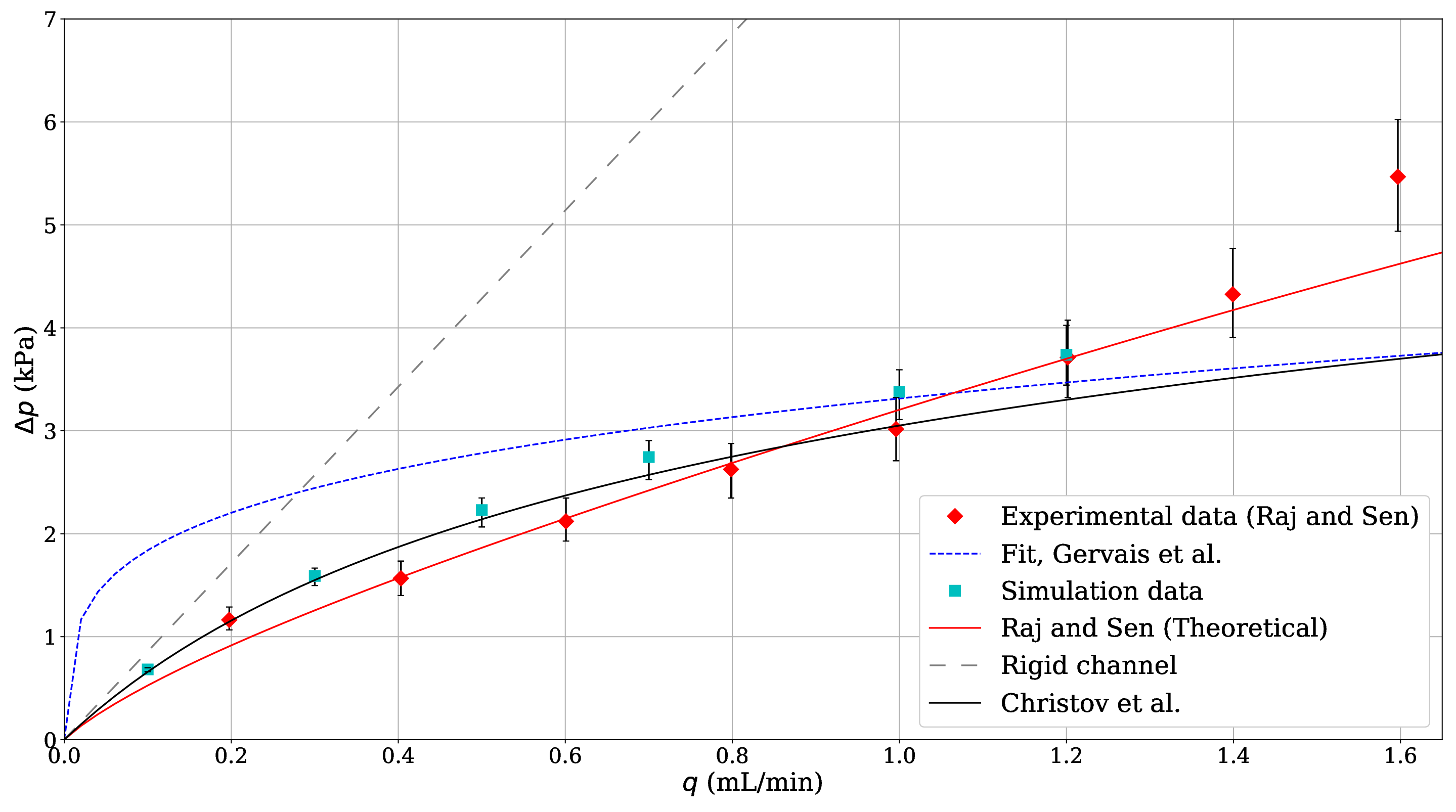}
\end{center}
\caption{$q$ versus $\Delta p$ for data set RS1. Error bars on simulation data represent a $\pm$0.2 MPa variation in the Young's modulus of PDMS. Error bars on the experimental data represent experimental uncertainty as reported in \cite{RS16}. A nonlinear least squares fit (modified as described in the text) gives $\alpha=48.5$ (see table~\ref{tb:params3}) to be used in equation~\eref{eq:q_p_gervais_dim}. The theoretical predictions from Christov et al.~\cite{CCSS17} and Raj and Sen~\cite{RS16} are given by equations~\eref{eq:q_p_icc_dim} and \eref{eq:q_p_rs_dim}, respectively.}
\label{fig:q-dp_RS1}
\end{figure*}

Figures \ref{fig:q-dp_OZ4}, \ref{fig:q-dp_OZ5} and \ref{fig:q-dp_RS1} show the variation of the flow rate $q$ with the pressure drop $\Delta p$ for the OZ4, OZ5 and RS1 data sets, respectively. Specifically, digitized experimental data from \cite{OYE13,RS16} is compared against our direct numerical simulations and the various theoretical curves. To highlight the effect of fluid--structure interactions, the rigid-channel lubrication theory relation $\Delta p = 12\mu \ell q/(h_0^3w)\left(1 - 0.630\frac{h_0}{w}\right)^{-1}$ (taking into account the leading-order side walls drag \cite[\S3.4.2]{B08}) is shown as a dashed line in all plots.

For the case of OZ4 in \fref{fig:q-dp_OZ4}, we observe very good agreement between experiment, simulation and theory. Our simulations and theoretical prediction fall within the estimated error bars of the experimental data. Additionally, the theoretical relationship between $q$ and $\Delta p$ given by equation~\eref{eq:q_p_icc_dim} agrees to a few percent with the simulation results. This agreement highlights that the OZ4 experimental data set's conditions are well within the thin-plate bending theory proposed in \cite{CCSS17} and summarized in section~\ref{sec:bending}. The fit of Gervais et al.~\cite{GEGJ06} given by equation~\eref{eq:q_p_gervais_dim} is also shown. The fit shows better agreement with the experimental data but this is always going to be the case because $\alpha$ is fit under this precise requirement.

For the case of OZ5 in \fref{fig:q-dp_OZ5}, we observe a large discrepancy between the experimental data and the simulation results. We conjecture that this is related to the ``parasitic pressure drop'' discussed in \cite{OYE13} as result of non-ideal experimental conditions. This effect is difficult to control experimentally. At first look, it may seem that the experimental data is shifted up by a constant $\Delta p_\mathrm{parasitic}$. However, because of fluid--structure interactions, $\Delta p_\mathrm{parasitic}$ likely depends on $q$, and this dependence has not been determined.\footnote{It is worthwhile to note that our simulation results agree well with the ``simple fits'' discussed in  \cite{OYE13}, in which the pressure drop is approximated as $\Delta p \approx q\int_0^\ell r(z)\,dz$. The hydraulic resistance $r(z)$ is calculated based on an approximation of the expression for a rectangular channel but with $h_0$ replaced by the appropriate $h(z)$ calculated from the experimental bulge-test constitutive curves measured in  \cite{OYE13}. This suggests that our simulations and theoretical predictions are consistent with the fluid--structure interactions characterized in \cite{OYE13}, and the discrepancy seen in figure~\ref{fig:q-dp_OZ5} is due to measurement uncertainty in $\Delta p$.} Nevertheless, we observe that our simulation data follows the trend of the experiments. Moreover, a good agreement can be observed with the predictions of the thick-plate expression [given by equation~\eref{eq:q_p_thick_dim}]. In this case, even though the fit given by equation~\eref{eq:q_p_gervais_dim} shows reasonable agreement with the experimental data (as is expected from a fit), the curvature is incorrect because the best-fit value of $\alpha$ is now negative! Since $\alpha$ is meant to represent some grouping of physical parameters, it cannot be negative. On physical grounds alone, fluid--structure interactions increase the local cross-sectional area of the channel, which must decrease the velocity, therefore decrease the viscous losses and, hence, \emph{decrease} the pressure drop for a fixed flow rate. This highlight a significant problems with using a fitting expression rather than a flow rate--pressure drop curve derived from first principles.

For the case of RS1 in \fref{fig:q-dp_RS1}, the pressure drop is \emph{not} the full pressure drop across the microchannel as was the case for OZ4 and OZ5. Instead, $\Delta p$ is calculated as $\Delta p = p(z=12\;\mathrm{mm}) - p(z=24\;\mathrm{mm})$ to be consistent with the experiments in \cite{RS16}. {For this reason, the fit based on equation~\eref{eq:q_p_gervais_dim} has to be calculated differently than for the previous two experimental data sets. First equation~\eref{eq:q_p_gervais_dim} is solved for $p(z)$. Second, the latter is used to develp an expression for the partial pressure drop $p(z=12\;\mathrm{mm}) - p(z=24\;\mathrm{mm})$. Third, this partial pressure drop expression is fit to the data to obtain the numerical value of $\alpha$.} Unlike figures~\ref{fig:q-dp_OZ4} and \ref{fig:q-dp_OZ5}, in \fref{fig:q-dp_RS1} we also show the theoretical prediction of Raj and Sen \cite{RS16}, namely equation~\eref{eq:q_p_rs_dim} (the compliance parameter $f_\mathrm{p}$ is given in table~\ref{tb:params3}). Although there is some visual discrepancy, our theoretical prediction passes within the error bars of most experimental data points and agrees very well with the simulation results up to $q\approx 1$ mL/min. A clear departure from the theoretical prediction (based on plate bending) can be observed for $q \gtrsim 1$ mL/min, and the theoretical prediction of Raj and Sen \cite{RS16} [i.e., equation~\eref{eq:q_p_rs_dim}] now matches the simulations. Thus, we conjecture that the elastic response is crossing over from the bending-dominated to a stretching-dominated regime at $q\approx 1$ mL/min in this data set. {Finally, we note that even with a best-fit $\alpha$ value, equation~\eref{eq:q_p_gervais_dim} does \emph{not} provide an accurate quantitative description of the RS1 flow rate--pressure drop data.}

\subsection{Pressure profile within the channel}
\label{sec:results_pressure}
Having established good agreement between experimental flow rate--pressure drop data, the corresponding theoretical predictions and our simulations, we now turn to the pressure profiles within the microchannel. Under the lubrication approximation, the pressure is expected to vary significantly only in the stream-wise $z$-direction. 

After introducing the dimensionless variables for the flow-rate-controlled regime, i.e., $Q=q/q=1$, $P = {p}/{\Delta p_\mathrm{c}}$,  where $\Delta p_\mathrm{c} = {\mu \ell} q/(h_0^3w)$ is the characteristic viscous pressure drop, and $Z={z}/{\ell}$, then the dimensionless pressure $P$ at a given dimensionless stream-wise location $Z$ can be found by inverting the algebraic equation (see \cite{CCSS17}):
\begin{eqnarray}
12(1-Z) = P(Z)\Bigg[ 1 &+ \frac{\tilde\beta}{20}P(Z) + \frac{\tilde\beta^2}{630} P(Z)^2 \nonumber\\ 
&+ \frac{\tilde\beta^3}{48\,048} P(Z)^3 \Bigg],
\label{eq:theory_P_Z}
\end{eqnarray}
where $\tilde{\beta} := \beta/24 ={w^4\Delta p}/(24 B h_0)$ is, as before, the dimensionless parameter quantifying the compliance of the top wall. Note that equation~\eref{eq:theory_P_Z} is just the rearranged and dimensionless version of equation~\eref{eq:q_p_icc_dim}. Equation~\eref{eq:q_p_thick_dim} can also be made dimensionless in the same manner:
\begin{eqnarray}
12(1-Z) = &P(Z)\Bigg[ 1 + \left(\frac{1}{20} + 24 f_1\right) \tilde\beta P(Z)  \nonumber\\ 
&+ \left(\frac{1}{630} + 576 f_2 \right) \tilde\beta^2 P(Z)^2  \nonumber\\
&+ \left( \frac{1}{48\,048} + 13\,824 f_3\right) \tilde\beta^3 P(Z)^3 \Bigg],
\label{eq:theory_P_Z_thick}
\end{eqnarray}
where the (dimensionless) functions $f_1(t/w)$, $f_2(t/w)$ and $f_3(t/w)$ are given by equations~\eref{eq:f1}, \eref{eq:f2} and \eref{eq:f3}, respectively.

From our numerical simulations, we can compute $P(X,Y,Z)$. To compare the numerical results to equations~\eref{eq:theory_P_Z} and \eref{eq:theory_P_Z_thick}, we evaluate the pressure from the simulations at $X=Y=0$ without loss of generality, since it is expected that the pressure does not vary significantly across the cross-section (which we have verified is indeed the case).

\begin{figure}[t]
\begin{center}
\includegraphics[width=\columnwidth]{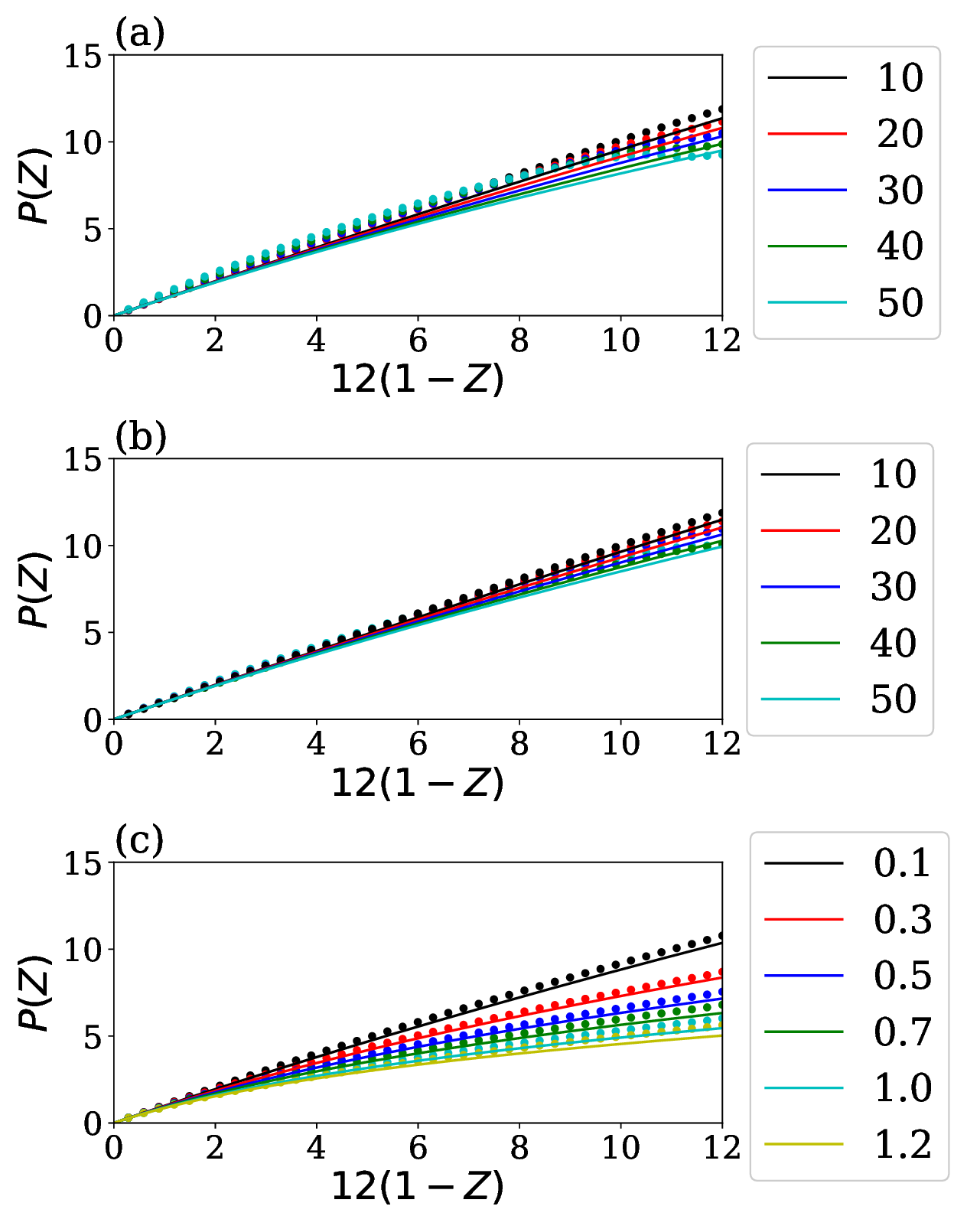}
\end{center}
\caption{Variation of the dimensionless pressure $P(Z)$ as a function of $12(1-Z)$ from simulations corresponding to the data sets (a) OZ4 , (b) OZ5  and (c) RS1. Dots indicate the simulation results, while solid curves are the theoretical prediction \eref{eq:theory_P_Z} (a,c) or \eref{eq:theory_P_Z_thick} (b). The legends in each panel indicate the \emph{dimensional} flow rate $q$ in mL/min. Since the plots show dimensionless quantities, increasing $q$ corresponds to increasing $\beta$ (see tables~\ref{tb:params1}--\ref{tb:params3}).}
\label{fig:P_vs_1-Z}
\end{figure}

\Fref{fig:P_vs_1-Z} shows the variation of the dimensionless pressure $P(0,0,Z)$, or ``$P(Z)$'' for short, as a function of $12(1-Z)$ for all the three microchannels. Different sets of curves correspond to different flow rates in the experiments/simulations. Under our nondimensionalization above, increasing the flow rate corresponds to increasing $\tilde\beta$ (i.e., the effects of fluid--structure interaction are more pronounced, recall that $\tilde\beta=0$ gives a rigid channel). Therefore, figure \ref{fig:P_vs_1-Z} can also be interpreted as showing the effect of varying $\tilde{\beta}$ with the ``lower'' curves and data corresponding to higher $\tilde{\beta}$.

As the three panels of figure \ref{fig:P_vs_1-Z} show, there is good agreement between the theoretical predictions and the simulation data  for all flow rates in RS1 and the lower flow rates for OZ4 and OZ5. The range of $\tilde{\beta}$ values is smallest for OZ5 [figure~\ref{fig:P_vs_1-Z}(b)], while it is largest for RS1 [figure~\ref{fig:P_vs_1-Z}(c)]. In general, since equations~\eref{eq:theory_P_Z} and \eref{eq:theory_P_Z_thick} are obtained using the lubrication approximation, we expect some disagreement for higher values of flow rates (i.e., at higher Reynolds numbers). However, even when there is a nontrivial difference between the theory and simulation data in figure~\ref{fig:P_vs_1-Z}, this discrepancy does not seem to affect the agreement between the computed and predicted flow rate--pressure drop relations (figures~\ref{fig:q-dp_OZ4}--\ref{fig:q-dp_RS1}), at least for the range of $\tilde{\beta}$ achieved here. 

Furthermore, some disagreement can be seen between the theoretical and simulation curves, for the case of OZ4 and OZ5, near the inlet of the microchannel [i.e., for $12(1-Z)\approx 1$]. This can be attributed to clamping of the top wall at the microchannel's inlet and outlet planes, which does not feature in the leading-order asymptotic solution for the displacement [i.e., equations~\eref{eq:u_bend_dim} and \eref{eq:u_bend_thick}]. In \cite{CCSS17}, based on the Kirchhoff--Love thin-plate equation, it was argued that clamping the top wall at the microchannel's inlet and outlet planes results in boundary-layer type corrections to $u$ localized within regions that are a fraction of the channel length $\ell$, on the order of $\epsilon/\delta = w/\ell$. In other words, these flow-wise boundary layers are on the order of the channel width. For OZ4 and OZ5 [figure~\ref{fig:P_vs_1-Z}(a,b)], $w/\ell=0.11$, hence we expect the effects of clamping to be localized to $12(1-Z) \gtrsim 10.68$. Meanwhile for RS1 [figure~\ref{fig:P_vs_1-Z}(c)], $w/\ell=0.014$, hence we expect the effects of clamping to be localized to $12(1-Z) \gtrsim 11.8$. Both of these estimates are clearly consistent with the simulation results shown in figure~\ref{fig:P_vs_1-Z}.

\subsection{Maximum displacement of the top wall}

The maximum channel height in any cross-section perpendicular to the flow is $h_\mathrm{max} = h_0 + u_\mathrm{max}$. For a thin plate, $u_\mathrm{max}$ is found from equation~\eref{eq:u_bend_dim} (dimensional) or equation~\eref{eq:u_bend_dimless} (dimensionless), from which the maximum channel height follows:
\begin{equation}
h_\mathrm{max}(z) = h_0 + \frac{w^4}{384 B} p(z) .
\label{eq:h_max}
\end{equation}
As before, $p(z)$ is found (in this model) by inverting equation~\eref{eq:q_p_icc_dim} for a given $q$. For a thick plate, using equation~\eref{eq:u_max_thick}, the maximum channel height is
\begin{equation}
h_\mathrm{max}(z) = h_0 + \frac{w^4}{384 B} \left[ 1+\frac{8(t/w)^2}{\kappa(1-\nu_\mathrm{s})} \right] p(z),
\label{eq:h_max_thick}
\end{equation}
where $p(z)$ is now found by inverting equation~\eref{eq:q_p_thick_dim} for a given $q$. In this subsection, we benchmark the theoretical predictions given by equations~\eref{eq:h_max} and \eref{eq:h_max_thick} against our simulation results and the experimental data from the literature.

\begin{figure}
\begin{center}
\includegraphics[width=\columnwidth]{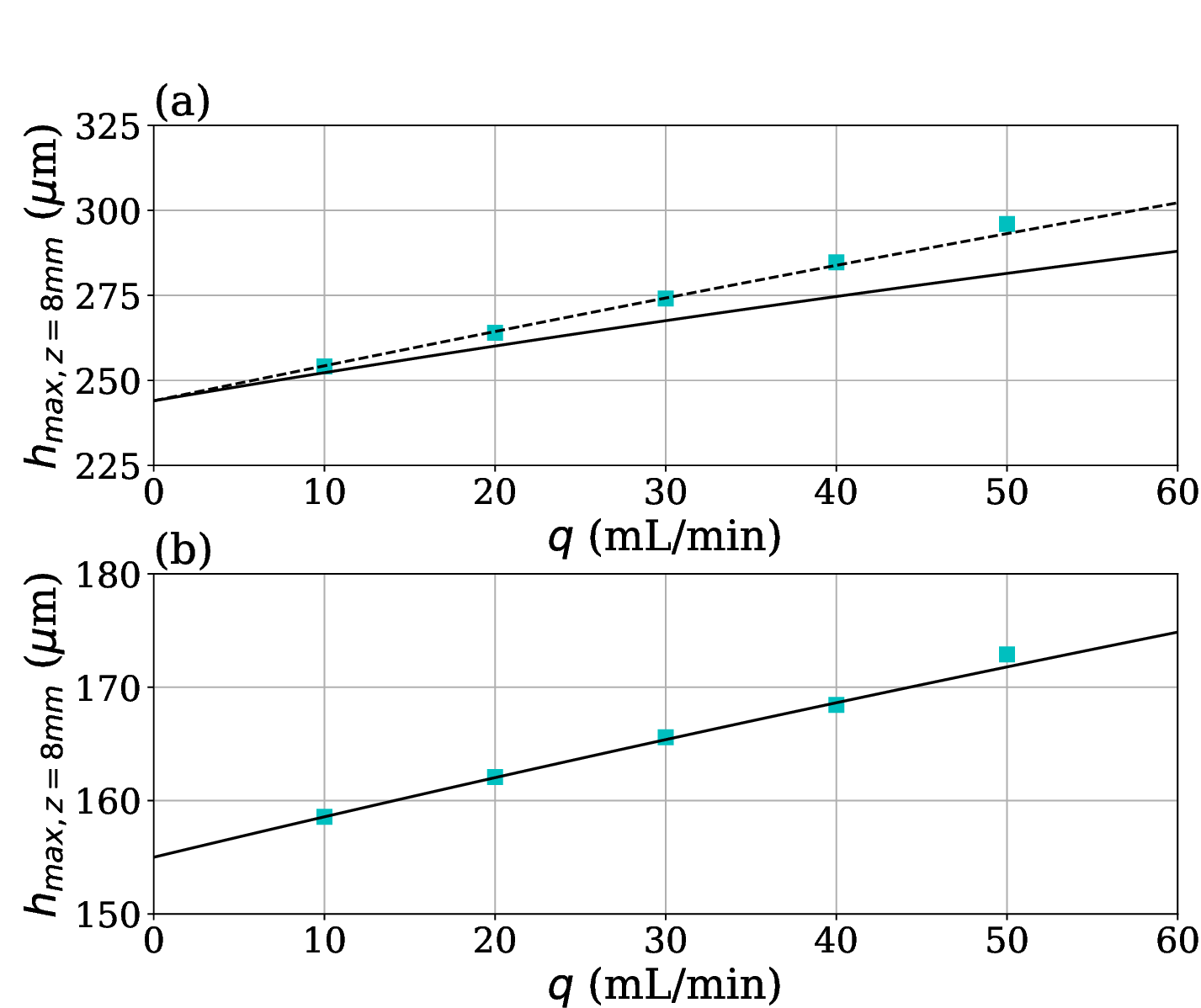} 
\end{center}
\caption{Variation of maximum channel height $h_\mathrm{max}$ at $z=8$ mm with the flow rate $q$ for the (a) OZ4 and (b) OZ5 simulations. In (a), the theoretical expressions from \cite{CCSS17} [given by equation~\eref{eq:h_max}] and also equation \eref{eq:h_max_thick} with $\kappa=1$ are shown as a solid and a dashed curve, respectively. In (b), only equation \eref{eq:h_max_thick} with $\kappa=1$ is shown as a solid curve. Data points correspond to the numerical simulations.}
\label{fig:OZ_h_max_vs_q}
\end{figure}

Ozsun et al.~\cite{OYE13} reported absolute maximum (i.e., maximum over both the span-wise $x$-direction and flow-wise $z$-direction) displacements of 86 and 25 $\mu$m for the highest flow rates for the OZ4 and OZ5 data sets, respectively. Our simulations compute a maximum displacement of 75.27 $\mu$m for OZ4 and 28.35 $\mu$m for OZ5 (both for $E = 1.6$ MPa), which are comparable to the experiments. Furthermore, in figure~\ref{fig:OZ_h_max_vs_q}, we compare our computational results for the maximum channel height in the span-wise direction at $z=8$ mm with the theoretical ones, across a range of flow rates.

Figure~\ref{fig:OZ_h_max_vs_q}(b) shows excellent agreement between the thick-plate prediction in equation~\eref{eq:h_max_thick} and the simulation data for OZ5. Meanwhile figure~\ref{fig:OZ_h_max_vs_q}(a) shows an acceptable agreement with the thin-plate prediction in equation~\eref{eq:h_max} for OZ4, especially for low flow rates. Since, even for OZ4 data set the top wall is not very thin, the thick-plate expression \eref{eq:h_max_thick} can be applied here as well (dashed curve in the plot). It is evident that this remedies most of the disagreement between the predicted maximum channel height at $z=8$ mm and the simulation.

\begin{figure}
\begin{center}
\includegraphics[width=\columnwidth]{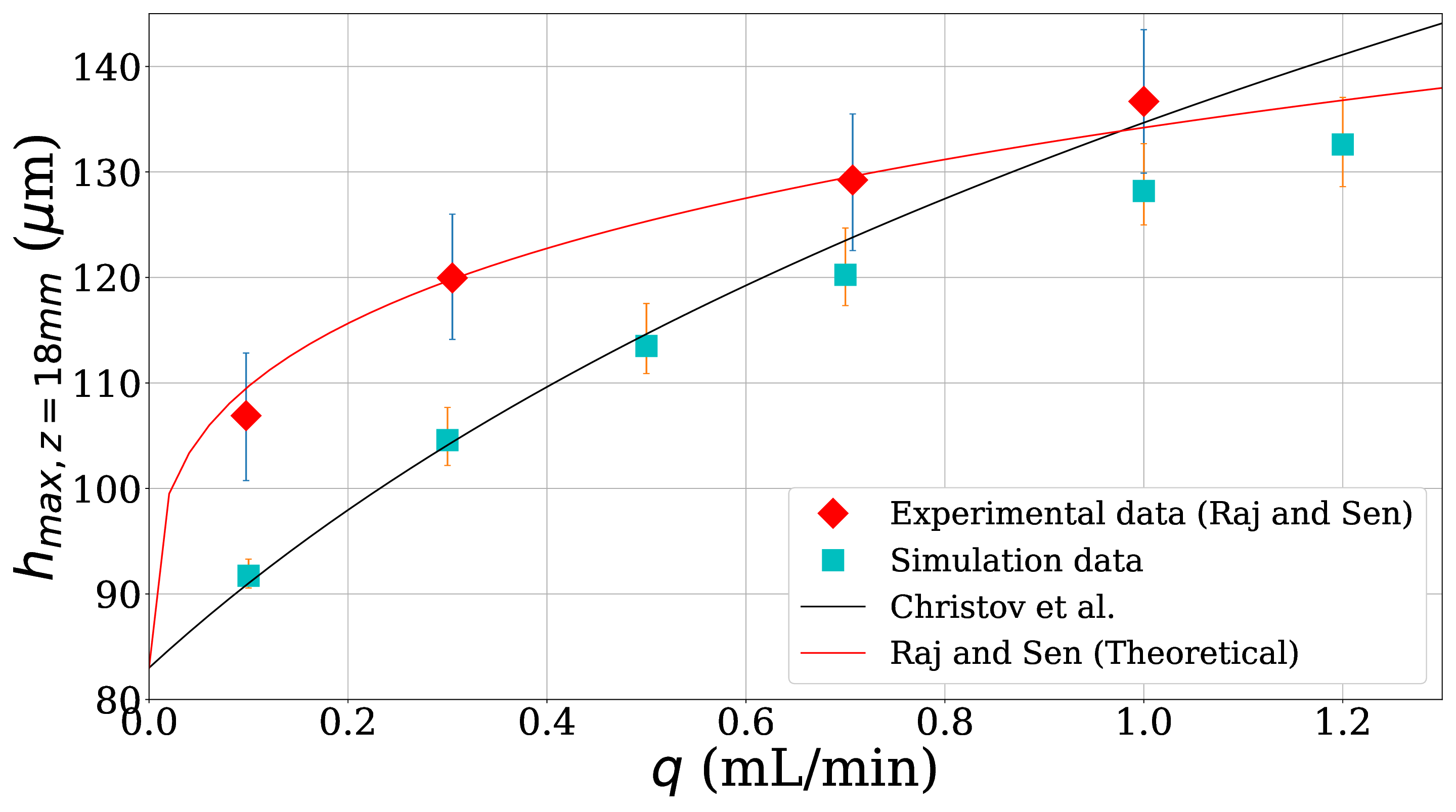} 
\end{center}
\caption{Variation of maximum channel height $h_\mathrm{max}$ at $z=18$ mm with the flow rate $q$ for the RS1 data set and simulations. The error bars on the experimental data indicate the uncertainty estimated in \cite{RS16}. Error bars on the simulation data indicate variation in the Young's modulus. The theoretical expressions from \cite{RS16} and \cite{CCSS17} [the latter given by equation~\eref{eq:h_max}] are shown as solid curves.}
\label{fig:RS1_h_max_vs_q}
\end{figure}

Raj and Sen \cite{RS16} obtained further displacement measurements than those reported by Ozsun et al.~\cite{OYE13}. Hence, in figure~\ref{fig:RS1_h_max_vs_q}, we compare our theoretical prediction and simulation results for the maximum channel height in the span-wise direction at $z=18$ mm (downstream of the channel inlet) with the RS1 experimental data, across a range of flow rates.

Consistent with the discussion in section~\ref{sec:results_qdp}, figure~\ref{fig:RS1_h_max_vs_q} shows a \emph{cross-over} from a plate-bending to a membrane-stretching regime near $q \approx 1$ mL/min. Our simulations show very good agreement with the thin-plate-bending expression \eref{eq:h_max} at low flow rates and somewhat better agreement with the stretching-based result from \cite{RS16}, i.e., $h_\mathrm{max} = h_0 + u_\mathrm{max}$ with  $u_\mathrm{max}$ given by equation \eref{eq:u_max_rs}. However, our theory and simulations do not agree well with the lowest two flow rate measurements from \cite{RS16}. We conjecture that this is due to significant experimental measurement uncertainty since our simulations clearly show that the low-$q$ regime for the RS1 data set is bending dominated. 

\begin{figure*}
\begin{center}
\includegraphics[width=0.9\textwidth]{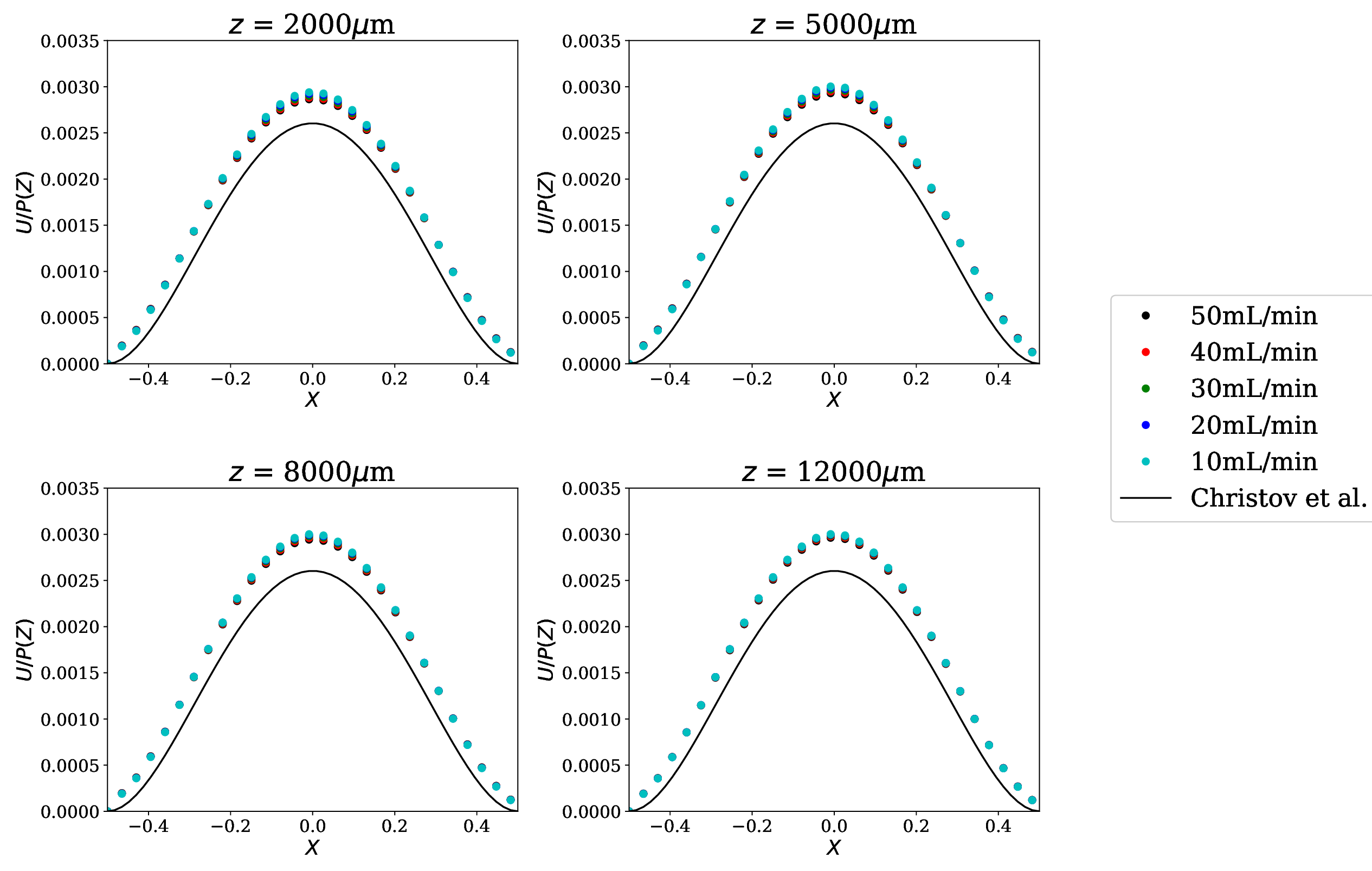}
\end{center}
\caption{Self-similar collapse (for different flow rates $q$ from 10 to 50 mL/min) of the cross-sectional deformation profile, $U(X,Z)/P(Z)$ versus $X$ for OZ4. Each of the four panels corresponds to a different flow-wise location, $z$. The solid curves correspond to the theoretical profile given in equation~\eref{eq:u_bend_dimless}. Note that the simulation data collapses very well for different $q$, thus many of the symbols overlap.}
\label{fig:U_by_P_vs_X_OZ4}
\end{figure*}

\begin{figure*}
\begin{center}
\includegraphics[width=0.9\textwidth]{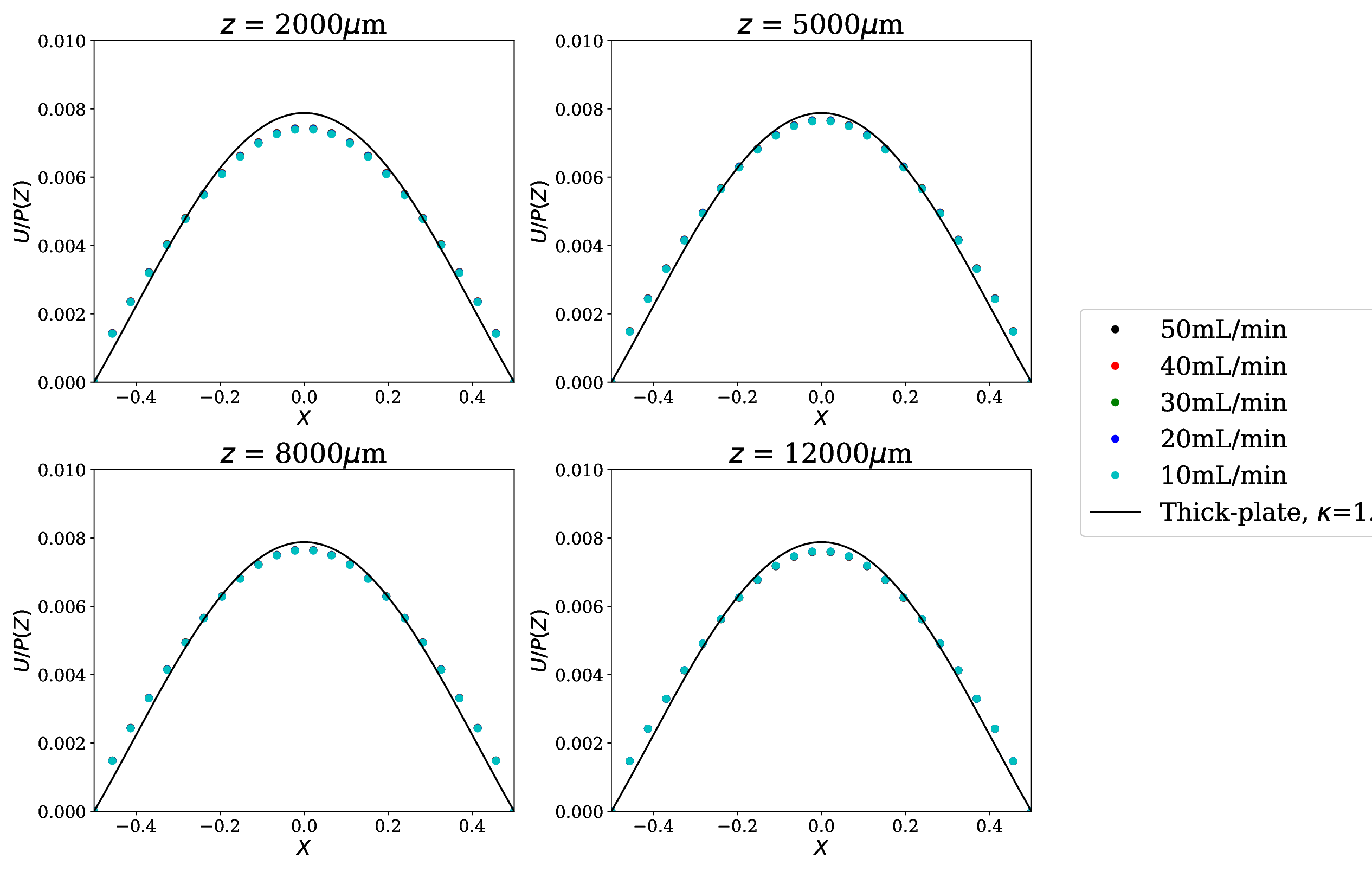}
\end{center}
\caption{Self-similar collapse (for different flow rates $q$ from 10 to 50 mL/min) of the cross-sectional deformation profile, $U(X,Z)/P(Z)$ versus $X$ for OZ5. Each of the four panels corresponds to a different flow-wise location, $z$. The solid curves correspond to the theoretical profile given in equation~\eref{eq:u_bend_thick_dimless}. Note that the simulation data collapses very well for different $q$, thus many of the symbols overlap.}
\label{fig:U_by_P_vs_X_OZ5}
\end{figure*}

\begin{figure*}
\begin{center}
\includegraphics[width=0.9\textwidth]{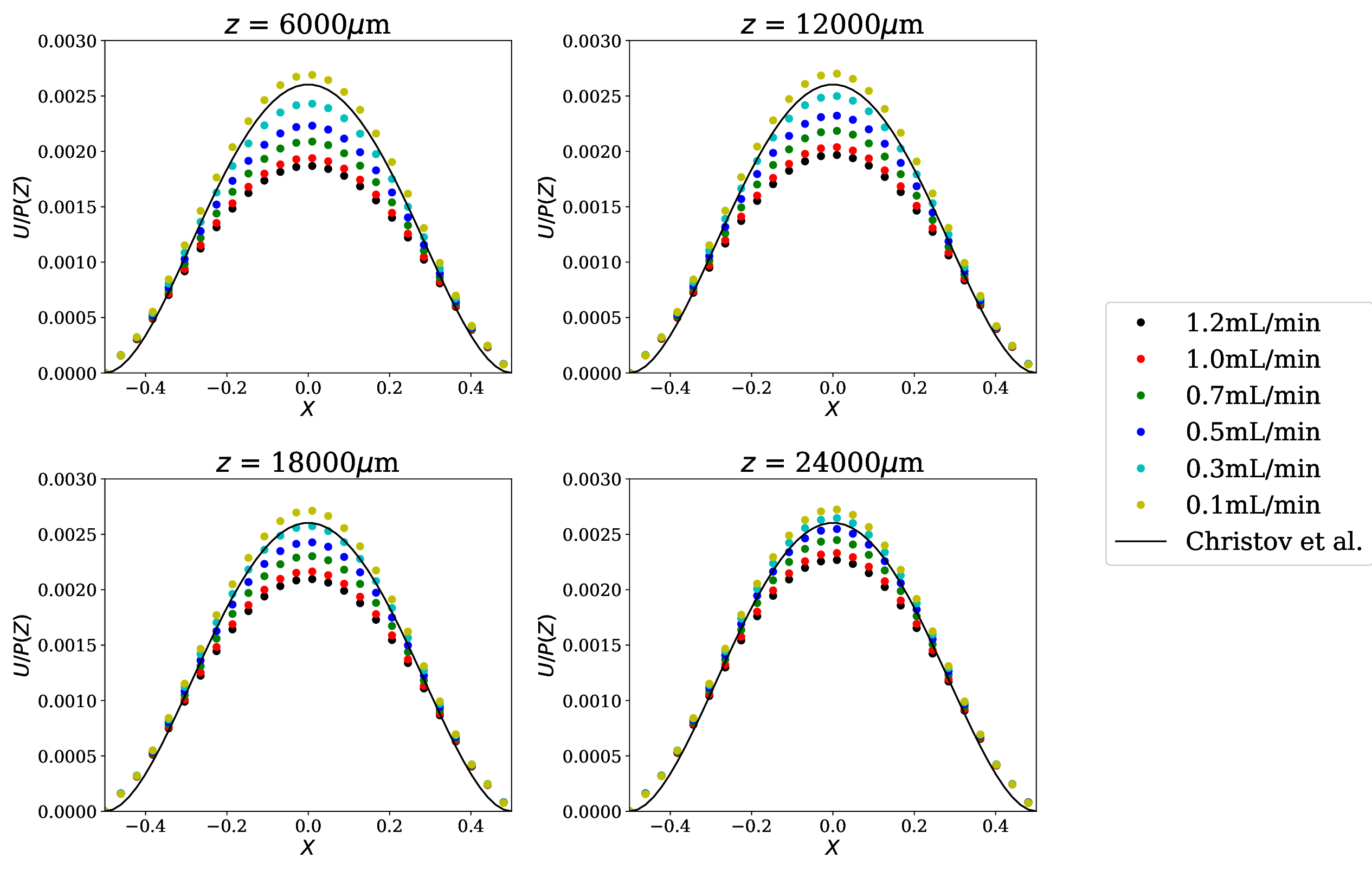}
\end{center}
\caption{Re-scaled cross-sectional deformation profiles (for different flow rates $q$ from 0.1 to 1.2 mL/min), $U(X,Z)/P(Z)$ versus $X$ for RS1. Each of the four panels corresponds to a different flow-wise location, $z$. The solid curves correspond to the theoretical profile given in equation~\eref{eq:u_bend_dimless}.}
\label{fig:U_by_P_vs_X_RS1}
\end{figure*}

\subsection{Deformation profiles}
\label{sec:results_deformation}

Fluorescence microscopy has been used as a cost-effective method to detect deformation of microchannels from their rectangular moulding \cite{HUZK09,KRO14,RS16,RDC17}. Even tough this technique has allowed for the measurement of the maximal cross-sectional displacement of a microchannel, the inherent uncertainties and noisy measurements produced by this technique do not provide a reliable quantitative displacement profile across the microchannel's cross-section. Only a few studies have employed the more accurate  confocal microscopy \cite{GEGJ06,NNBR17}. On the other hand, our mathematical models do provide an analytical expression for the cross-sectional displacement profile, and we can easily compare this result to our 3D direct numerical simulations. This is the purpose of this subsection.

Specifically, under the lubrication approximation, we expect the dimensionless $U(X,Z)/P(Z)$ curves to collapse (specifically, to become independent of $Z$ and $Q$) as suggested by equation \eref{eq:u_bend_dimless} (thin-plate) or equation~\eref{eq:u_bend_thick_dimless} (thick-plate). Figures \ref{fig:U_by_P_vs_X_OZ4}, \ref{fig:U_by_P_vs_X_OZ5} and \ref{fig:U_by_P_vs_X_RS1} show the the rescaled deformation profiles for the OZ4, OZ5 and RS1 data sets, respectively. In figures \ref{fig:U_by_P_vs_X_OZ4}--\ref{fig:U_by_P_vs_X_RS1}, each of the four panels shows a particular flow-wise location (i.e., fixed $z$) along the microchannel, in which the displacement profiles for all flow rates have been made dimensionless and plotted together. For the OZ4 and OZ5 data sets, we observe excellent self-similar collapse (with respect to both flow rate and $z$ location) and fairly good agreement with the analytical predictions from equation~\eref{eq:u_bend_dimless} (thin-plate) or equation~ \eref{eq:u_bend_thick_dimless} (thick-plate), respectively. Note that, from table~\ref{tb:dims}, we can calculate the plate aspect ratios for the OZ4 and OZ5 data sets to be $t/w=0.12$ and $0.36$, hence we are to expect an agreement with equation \eref{eq:u_bend_dimless} (thin-plate) for the OZ4 case, and an agreement with equation \eref{eq:u_bend_thick_dimless} (thick-plate) for the OZ5 case. 


For the RS1 data set however, we do not see a collapse of the $U(X,Z)/P(Z)$ profiles. In order to ensure that this is not a flow-development effect,\footnote{Recall that, in this case, to match the experimental conditions, we impose a uniform velocity profile at the inlet in our simulations.} we re-ran several cases with a fully developed inlet profile. We did not observe a significant change in the ``quality'' of the collapse. We attribute the observed spread in the curves in figure~\ref{fig:U_by_P_vs_X_RS1} to the fact that the elastic response is crossing over from bending-dominated at low flow rates to stretching-dominated at higher flow rates. Nevertheless, since $t/w=0.11\ll 1$, the low-flow-rate simulation data in figure~\ref{fig:U_by_P_vs_X_RS1} agrees well with the thin-plate-bending profile given in equation~\eref{eq:u_bend_dimless}.  

Finally, we note that a value of $\kappa = 1$ is chosen for the thick-plate displacement profile [equation~\eref{eq:u_bend_thick_dimless}] because, as Zheng \cite{Z06} has shown, convergence of the Mindlin plate model's solution to that the of the full elasticity equations (as $t/w\to 0^+$) requires that $\kappa=1$. Moreover, the traditional value $\kappa=5/6$ \cite{H01,GW01} is most relevant when attempting to derive Timoshenko's beam equation from those of a Mindlin plate, taking into account shear deformation \cite{C66}. In our asymptotic regime of $h_0 \ll w \ll \ell$, we do not expect shear in the $z$-direction to be significant, with each span-wise cross-section acting essentially as a beam \cite{CCSS17}.


\section{Conclusion}
\label{sec:conc}

In this work, we presented a detailed theoretical--computational study of the static response of long shallow microchannels. Specifically, the flow-induced shape deformation of microchannels of rectangular cross-section with a soft top wall was analysed. We adopted the theoretical approach of Christov et al.~\cite{CCSS17} to the coupled problem. Specifically, we derived expressions for the flow rate--pressure drop relation, the pressure profile in the microchannel and the cross-sectional top-wall displacement. Restricting to isotropic linearly elastic materials, we considered top walls that act both as thin and thick plates. 

The theoretical developments were benchmarked against full 3D two-way fluid--structure interaction (FSI) simulations using ANSYS$^\textsuperscript{\textregistered}$ Workbench, showing excellent agreement with the theory. This paves the way for using fitting-parameter-free mathematical expressions that do not require experimental characterization for model closure, such as equation~\eref{eq:q_p_thick_dim}, in the design of microfluidic systems. Additionally, our simulations, which are some of the few 3D FSI simulations of microchannels, were both validated (compared to three different types of experiments from the literature) and verified (grid-convergence was shown), in the terminology of Oberkampf and Trucano~\cite{OT02}. Moreover, these high-resolution numerical simulations provide details about the physics of microfluidic FSIs that are difficult (or impossible) to obtain experimentally, namely accurate cross-sectional displacement profiles.

The comparison of theory and simulations specifically confirmed several results, which we discussed follow from the modelling approach: (i) the $q$--$\Delta p$ relation is a quartic polynomial (i.e., not linear); (ii) the dimensionless cross-sectional displacement scaled by the dimensionless pressure, i.e., $U(X,Z)/P(Z)$, shows collapse for different flow rates and flow-wise locations, which confirms the prediction that the deformation in a span-wise cross-section and the flow-wise deformation are decoupled; (iii) several of the benchmark experimental systems for flow-induced deformation of microchannels are well described by a plate-bending-based theory. 

A remaining open question is the effect of stretching in a thin top wall. The RS1 data above appears to crossover from a plate-bending regime ($u/w\propto p/E$) at low flow rates into a membrane-stretching regime [$u/w\propto (p/E)^{1/3}$] at high flow rates. Ozsun et al.~\cite{OYE13} also observed such stretching-dominated deformations when using nanometer-thickness SiN membranes, which are beyond the scope of the present discussion. The crossover behaviour could be modelled by the F\"oppl--von K\'{a}rm\'{a}n equations \cite{TWK59,HKO09}, in which case, an exact solution for the deformation $u(x,z)$ is unlikely. Nevertheless, a flow rate--pressure drop curve could be constructed (using the ``recipe'' we outlined in section~\ref{sec:models}) and evaluated fully numerically. Since it appears most of the microchannel deformation data (except the S1--S3 sets in \cite{OYE13} and the high-flow-rate data points in \cite{RS16}) is not in the stretching-dominated regime, it remains to be determined whether such a calculation is worthwhile exploring. 

Future work would include considering transient problems, in particular due to their applications in stop-flow lithography \cite{DGPHD07}. Extending our analysis to non-Cartesian geometries is also of interest. Fluid--structure interactions in collapsible tubes have a long history \cite{P80,GJ04}, but the interest has been primarily on moderate and high Reynolds numbers with applications to the dynamics of blood vessels. Nevertheless, low-Reynolds-number flows also arise in biofluid mechanics \cite{F97} and have also garnered some attention due to applications to soft robotics \cite{EG14}.

\ack
We wish to thank the special issue co-editors Prof.\ Peichun Amy Tsai (University of Alberta) and Prof.\ Rodrigo Ledesma (Northumbria University) for their kind invitation to contribute. This research is, in part, supported by the U.S.\ National Science Foundation under grant No.\ CBET-1705637.


\setcounter{section}{1}
\appendix
\section{Hydrostatic bulge tests and the determination of material properties}


Ozsun et al.~\cite{OYE13} performed a series of non-invasive hydrostatic bulge tests of their microchannel systems. In these tests, the top wall was subjected to a uniform hydrostatic pressure by plugging the outlet port of the microchannel, while connecting a water column to the microchannel's inlet. The resultant pressure drop--wall deformation curves are representative of the elastic material's constitutive behaviour, and thus involve the material properties of the compliant walls.

Several researchers in the past have derived analytical expressions to determine the pressure--deformation relationship for a variety of bulge test configurations (see, e.g., \cite{SN92} and the references therein). To the best of our knowledge, most of these studies deal with thin membranes, which have significantly higher $w/t$ ratios (e.g., ${w}/{t} \simeq 80-100$). For the OZ4 and OZ5 model systems discussed above, the walls are geometrically closer to a plate than a membrane: ${w}/{t} = 8.5$ and $2.81$, respectively. Finite element method (FEM) simulations have shown that expressions obtained from membrane theory are found to be inadequate in predicting the characteristics of thin and thick plates \cite{J08}. Thus, to get an accurate estimate of the Young's modulus for the two chosen compliant microchannels of Ozsun et al.~\cite{OYE13}, we carried out our own FEM simulations of the hydrostatic bulge test using ANSYS$^\textsuperscript{\textregistered}$ Workbench. 

\subsection{Geometry}

\begin{figure}[ht]
	\begin{center}
		\includegraphics[width=\columnwidth]{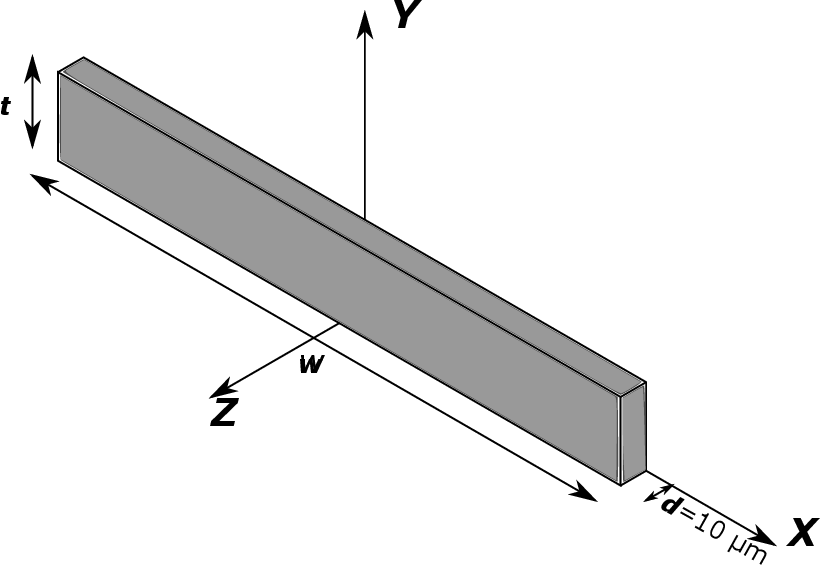}
	\end{center}
	\caption{Computational domain for bulge test simulations.}
	\label{fig:Geometry_bulge}
\end{figure}

From the channel rendering of the bulge test shown in \cite{OYE13}, it can be seen that the maximum deformation occurs along the microchannel's midplane, and the region of maximum deformation spans a significant portion of the central region of the channel's flow-wise extent. In order to simulate the wall deflection behaviour in this region, a thin (axial extent of $d = 10$ $\mu$m) strip was considered, as shown in \fref{fig:Geometry_bulge}. As before, $w$ and $t$ are the channel's width and the tio wall's thickness, respectively, and their values are given in table~\ref{tb:dims}.

From the experimental bulge test data, it was observed that the top wall in OZ4 has an initial non-zero deflection. Using a linear fit through the first three experimental data points, this initial height was estimated to be $\approx 8.38$ $\mu$m. However, no information has been provided in \cite{OYE13} regarding the exact value of this initial deformation or the initial shape of the channel. Thus, in our flow and bulge test simulations, we have modelled the top wall as being flat initially. We take into account the initial deformation by adding our estimate of 8.38 $\mu$m to the simulation values of the maximum deformation.

\subsection{Results}

\begin{figure}[ht]
	\begin{center}
	\includegraphics[width=\columnwidth]{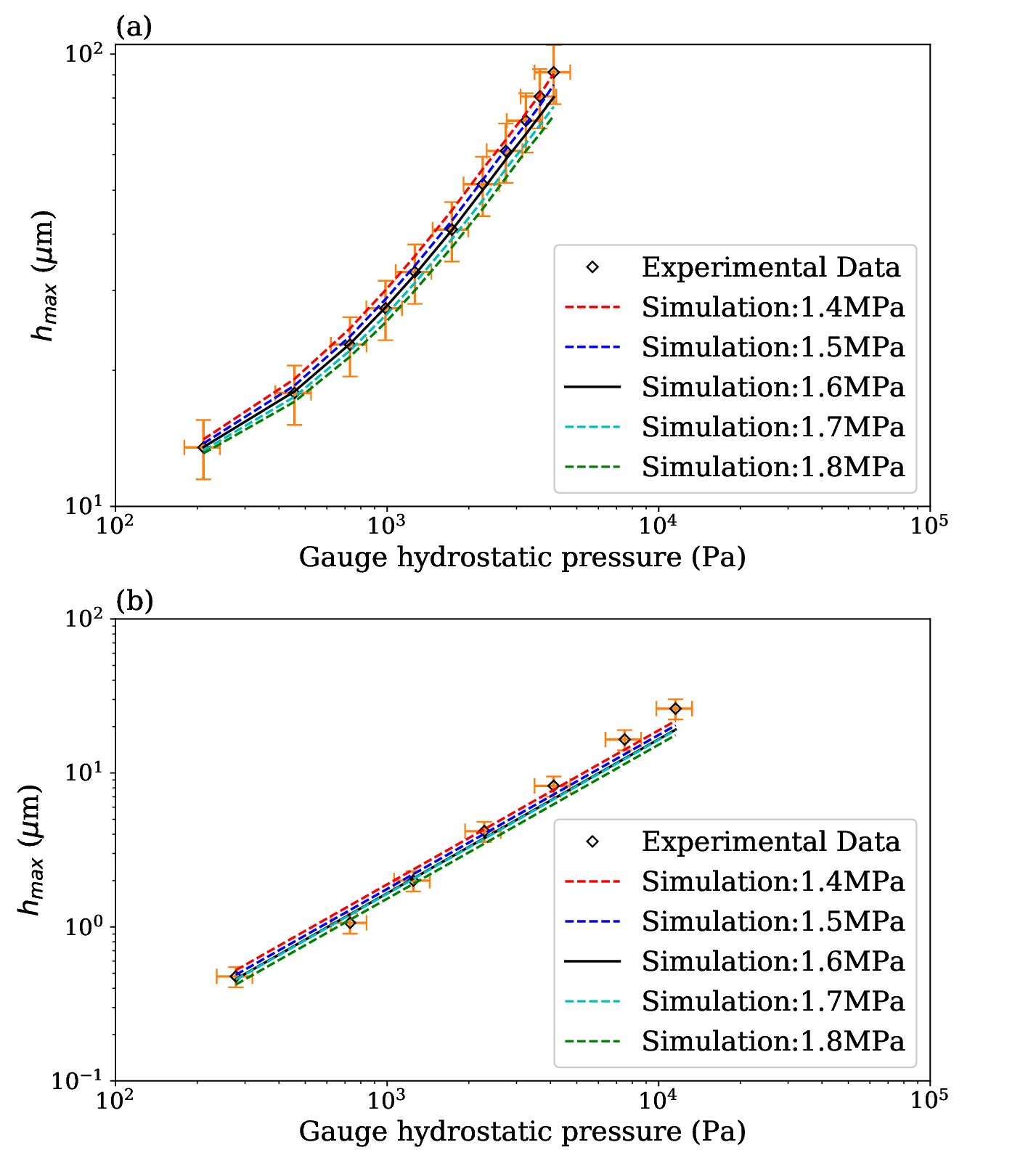}
	\end{center}
	\caption{Hydrostatic bulge test simulation results for the data sets (a) OZ4 and (b) OZ5. Error bars on the experimental data are based on the extent of the symbol sizes in \cite{OYE13}. A range of Young's moduli was considered, with the $E=1.6$ MPa being the base value used in the present study.}
	\label{fig:Bulge_Test}
\end{figure}

The plots shown in \fref{fig:Bulge_Test} compare the maximum channel height from the experiments of Ozsun et al.~\cite{OYE13} and our simulations, for the cases we denoted as (a) OZ4 and (b) OZ5. As can be seen in the figure, our simulations agrees with the experiments (within the error bars reported in \cite{OYE13}) for a range of Young's moduli. Although this agreement suggest that the range of Young's moduli from $1.4$ MPa to $1.8$ MPa is consistent with the hydrostatic bulge test data from \cite{OYE13}, the hydrostatic bulge test show that the mean value of $E=1.6$ MPa is a reliable baseline estimate. Therefore, we have used this value in our simulations discussed in the main text above.

\section*{References}
\bibliographystyle{iopart-num}
\bibliography{deformable_microchannels.bib}

\end{document}